\documentclass[10pt, oneside]{article}
\usepackage{geometry,color}
\usepackage{graphicx}
\usepackage{amsfonts,amsmath,amssymb,mathrsfs}

\newtheorem{theorem}{Theorem}[section]

\newtheorem{corollary}[theorem]{Corollary}

\newtheorem{definition}[theorem]{Definition}
\newtheorem{example}[theorem]{Example}

\newtheorem{lemma}[theorem]{Lemma}

\newtheorem{problem}[theorem]{Problem}
\newtheorem{proposition}[theorem]{Proposition}
\newtheorem{remark}[theorem]{Remark}
\numberwithin{equation}{section}

\newcommand{\beq}{\begin{equation}}
\newcommand{\eeq}{\end{equation}}
\newcommand{\beqn}{\begin{eqnarray}}
\newcommand{\eeqn}{\end{eqnarray}}
\newcommand{\bgn}{\begin{align}}
\newcommand{\egn}{\end{align}}
\newcommand{\bmrk}{\begin{remark}}
\newcommand{\emrk}{\end{remark}}
\newcommand{\beqm}{\begin{equation*}}
\newcommand{\eeqm}{\end{equation*}}
\newcommand{\beqnm}{\begin{eqnarray*}}
\newcommand{\eeqnm}{\end{eqnarray*}}
\newcommand{\bgnm}{\begin{align*}}
\newcommand{\egnm}{\end{align*}}

\newcommand{\bthm}{\begin{theorem}}
\newcommand{\ethm}{\end{theorem}}
\newcommand{\blem}{\begin{lemma}}
\newcommand{\elem}{\end{lemma}}
\newcommand{\beg}{\begin{example}}
\newcommand{\eeg}{\end{example}}
\newcommand{\bprop}{\begin{proposition}}
\newcommand{\eprop}{\end{proposition}}
\newcommand{\bplm}{\begin{problem}}
\newcommand{\eplm}{\end{problem}}

\title{On Realization Theory of Quantum Linear Systems \thanks{Corresponding author Guofeng Zhang. Tel. +852-2766-6936. 
Fax +852-2764-4382. Email: Guofeng.Zhang@polyu.edu.hk.}}

\author{John E. Gough\thanks{Institute of Mathematics and Physics, Aberystwyth
University, Ceredigion SY23 3BZ, Wales UK. (e-mail: jug@aber.ac.uk).} \and
Guofeng Zhang\thanks{Department of Applied Mathematics, The Hong Kong
Polytechnic University, Hong Kong, China. (e-mail:
Guofeng.Zhang@polyu.edu.hk).}}

\begin{document}
\maketitle

\begin{abstract}
The purpose of this paper is to study the realization theory of quantum linear systems. It is shown that for a general quantum linear system its controllability and observability are equivalent and they can be checked by means of a simple matrix rank condition. Based on controllability and observability a specific realization is proposed for general quantum linear systems in which an uncontrollable and unobservable subspace is identified. When restricted to the passive case, it is found that a realization is minimal if and only if it is Hurwitz stable. Computational methods are proposed to find the cardinality of minimal realizations of a quantum linear passive system.  It is found that the transfer function of a quantum linear passive system $G$ can be written as a fractional form in terms of a  matrix function $\Sigma$; moreover, $G$ is lossless bounded real if and only if $\Sigma$ is lossless positive real. A type of realization for multi-input-multi-output quantum linear passive systems is derived, which is closely related to its controllability and observability decomposition. Two realizations, namely the independent-oscillator realization and the chain-mode realization, are proposed for single-input-single-output quantum linear passive systems, and it is shown that under the assumption of minimal realization, the  independent-oscillator realization is unique, and these two realizations are related to the lossless positive real matrix function $\Sigma$. 

\end{abstract}

%%%%%%%%%%%%%%%%%%%%%%%%%%%%%%%%
%%%%%%%%%%%%%%%%%%%%%%%%%%%%%%%%
%%%%%%%%%%%%%%%%%%%%%%%%%%%%%%%%
\section{Introduction}\label{sec:introduction}

Linear systems and signals theory has been very useful in the analysis and
engineering of dynamical systems. Many fundamental notions have been
proposed to characterize dynamical systems from a control-theoretic point of view. For example, controllability
describes the ability of steering internal system states by
external input, observability refers to the possibility of reconstructing
the state-space trajectory of a dynamical system based on its external
input-output data. Based on controllability and observability, Kalman
canonical decomposition reveals the internal structure of a linear system.
This, in particular minimal realization as a very convenient and yet quite
natural assumption, is the basis of widely used model reduction methods such
as balanced truncation and optimal Hankel norm approximation. Moreover,
fundamental dissipation theory has been well established and has been proven
very effective in control systems design. All of these have been well
documented, see, e.g., \cite{KS72}; \cite{Wil72}; \cite{AV73}; \cite{Kailath80}; \cite{vdS96}; \cite{ZDG96}.

%Many properties of quantum linear systems have been explored from a
%systems and signals point of view, e.g., structures of quantum linear
%systems (\cite{Car93}, \cite{Gar93}, \cite{WM94}, \cite{YK03}, \cite{GJ09},\cite{GJN10a}, \cite{ZJ11}, \cite{ZLW+13}); notions like controllability, observability, and minimal realization (\cite{MP11}); quantum $H^\infty$
%control (\cite{JNP08}, \cite{Mabuchi08}, \cite{MP11}, \cite{ZJ11}); quantum
%linear quadratic Gaussian (LQG) control (\cite{DJ99}, \cite{WD05}, \cite{NJP09}, \cite{HHH+09}, \cite{ZJ11}, \cite{ZLHZ12}); risk-sensitive control (\cite{YB09}); dissipation (\cite{MP11a}, \cite{ZJ11}); model reduction (\cite%
%{IRP12}, \cite{HIN13}); synthesis (\cite{NJD09}); squeezing enhancement (\cite{WM94b}, \cite{GW09}, \cite{IYY+11}); single-photon pulse shaping (\cite%
%{Milburn08}, \cite{ZJ13}), optical switches (\cite{Mabuchi11}); waveform
%estimation (\cite{TWC11}), quantum memory (\cite{HCH13}), among others.

In recent years there has been a rapid growth in the study of quantum linear systems. Quantum linear systems and signals theory has been proven very effective in the study of many quantum systems including quantum optical systems, opto-mechanical systems, cavity quantum electro-magnetic dynamical systems, atomic ensembles and quantum memories, see, e.g., \cite{GZ00}; \cite{WM08}; \cite{WM10}; \cite{SvHM04}; \cite{ZCB+10}; \cite{MHP+11}; \cite{MJP+11}; \cite{Tian12}; \cite{ZLW+13}; \cite{HCH13}. Because of its analytical and computational advantages, the linear setting always serves as an essential starting point for development of a more general theory.

Controllability and observability of quantum linear passive systems have been discussed in \cite{MP11a}; these two properties are used to establish the complex-domain bounded real lemma (\cite[Theorem 6.5] {MP11a}) for quantum linear passive systems,  which is the basis of quantum $H^\infty$ coherent  feedback control of  quantum linear passive systems,  \cite{MP11b}. For a quantum linear passive system it is shown in \cite[Lemma 3.1]{GY13} that controllability is equivalent to observability; moreover, a minimal realization is necessarily Hurwitz stable, \cite[Lemma 3.2]{GY13}. In this paper we explore further controllability and observability of quantum linear systems. For general quantum linear systems (not necessarily passive), we show that controllability and observability are equivalent (Proposition \ref{prop:minimal_C_Omega}). Moreover, a simple matrix rank condition is established for checking controllability and observability. Base on this result, a realization of general quantum linear systems is proposed, in which the uncontrollable and unobservable subspace is identified (Theorem \ref{thm:DF}). Theorem \ref{thm:DF}  can be viewed as the complex-domain counterpart of Theorem 3.1 in \cite{NY13} in the real domain.  However,  it is can be easily seen from the proof of Lemma \ref{lem:ctrb_obsv} that the structure of the unitary transformation involved  is better revealed in the complex domain. Restricted to the passive case, we show that controllability, observability and Hurwitz stability are equivalent to each other (Lemma  \ref{thm:stab_obsr_cont}). Thus, the realization of a quantum linear passive system is minimal if and only if it is Hurwitz stable (Theorem \ref{thm:minimal_hurwitz}). We also derive formulas for calculating  the cardinality of minimal realizations of a given quantum linear passive system (Proposition \ref{prop:siso_minimal} for the single-input-single-output case and Proposition \ref{prop:mimo_minimal} for the multi-input-multi-output case). Finally we show how a given quantum linear system can be written as a fractional form in term of a matrix function $\Sigma$ (Proposition \ref{lem:G_Sigma_0}), and for the passive case show  that a quantum linear passive system $G$ is lossless bounded real if and only if the corresponding $\Sigma$  is lossless positive real (Theorem \ref{eq:G_Sigma_Z}).

The synthesis problem of quantum linear systems has been investigated in \cite{NJD09}, where they showed that a quantum linear system can always be realized by a cascade of one-degree-of-freedom harmonic oscillators with possible direct Hamiltonian couplings among them if necessary. Then in \cite{HIN10} a necessary and sufficient condition is derived for the realizability of quantum linear systems via pure cascading only. For the passive case, it is shown in \cite{IRP11} that, under certain conditions on the system matrices, a minimal quantum linear passive system can be realized by a cascade of one-degree-of-freedom harmonic oscillators. These restrictions were removed in \cite{HIN10} which proves that all quantum linear passive systems can be realised by pure cascading of one-degree-of-freedom harmonic oscillators. Model reduction of quantum linear systems has been studied in, e.g., \cite{IRP13}, and \cite{HIN13}.  In this paper we propose several realizations of quantum linear passive systems. For the multi-input-multi-output (MIMO) case we show that the proposed realization has a close relationship with controllability and observably of the quantum linear passive system (Theorem \ref{thm:MIMO}). In the single-input-single-output (SISO) case, we propose two realizations, namely the independent-oscillator realization and the chain-mode realization (Theorem \ref{prop:independent_oscillator} and Theorem \ref{lem:continued_fraction}), and finally we show that if the system is Hurwitz stable, these two realizations are related to the lossless positive real $\Sigma$ mentioned in the previous paragraph (Theorem \ref{thm:independent_mode_unique}).

The rest of the paper is organized as follows. Section \ref{sec:sys_general} studies general quantum linear systems; specifically, Subsection  \ref{subsec:sys_general} briefly reviews quantum linear systems,  Subsection \ref{subsec:stability_ctrb_obsv_general} investigates their controllability and observability, and Subsection \ref{subsec:tf_general} presents a fractional form for transfer functions of quantum linear systems. Section \ref{sec:passive} studies quantum linear passive systems, specifically,  Subsection \ref{subsec:passive} introduces quantum linear passive systems, Subsection \ref{sec:stability_minimality} investigates their Hurwitz stability, controllability and observability, Subsection \ref{subsec:minimal} studies minimal realizations of quantum linear passive systems, and Subsection \ref{sec:dissipation} proposes a fractional form for transfer functions of quantum linear passive systems. Section \ref{sec:representation} investigates realizations of quantum linear passive systems; specifically,  Subsection \ref{subset:MIMO} proposes a realization for MIMO quantum linear passive systems,  Subsections \ref{sec:IO} and \ref{subset:chain} propose an independent-oscillator realization and  a chain-mode realization for SISO quantum linear systems respectively, and Subsection \ref{sec:equivalence} discusses the uniqueness of the independent-oscillator realization. Section \ref{sec:conclusion} concludes this paper.

\emph{Notations.}~ $m$ is the number of input channels, and $n$ is the number of degrees of freedom of a
given quantum linear system, namely, the number of system oscillators. Given a column vector of complex numbers or operators
$x
=
[
\begin{array}{ccc}
x_1 & \cdots & x_k
\end{array}
]^T,
$
 define
$x^\#
=
[
\begin{array}{ccc}
x_1^\ast & \cdots & x_k^\ast
\end{array}
]^T
$,
where the asterisk $\ast$ indicates complex conjugation or Hilbert
space adjoint. Denote $x^\dag = (x^\#)^T$. Furthermore, define a column vector $\breve{x}$ to be
$
\breve{x}
=
[
\begin{array}{cc}
x^T & (x^\#)^T
\end{array}
]^T
$.  Let $I_k$ be an
identity matrix and $0_k$ a zero square matrix, both of dimension $k$.
Define $J_k=\mathrm{diag}(I_k,-I_k)$. Then for
a matrix $X\in\mathbb{C}^{2j\times 2k}$, define $X^\flat =J_k X^\dag J_j$. Given two constant matrices $U$, $V\in \mathbb{C}^{r\times k}$, define $\Delta(U,V) = [ U ~ V; V^\# ~ U^\#]$. Given two operators $A$ and $B$, their commutator is defined to be $[A,B]=AB-BA$. ``$\Longleftrightarrow$'' means if and only if. Finally, $\mathrm{Spec}(X)$ denotes the set of all \textit{distinct} eigenvalues of the matrix $X$, $\sigma ( X )$ denotes the diagonal matrix with diagonal entries being
the non-zero singular values of the matrix $X$, $\mathrm{Ker}\left( X\right)$ denotes the null space of the matrix $X$, and $\mathrm{Range}\left( X\right) $ denotes the space spanned by  the columns of the matrix $X$.

%%%%%%%%%%%%%%%%%%%%%%%%%%%%%%%%
%%%%%%%%%%%%%%%%%%%%%%%%%%%%%%%%
%%%%%%%%%%%%%%%%%%%%%%%%%%%%%%%%
\section{Quantum linear systems} \label{sec:sys_general}
We first introduce quantum linear systems in Subsection \ref{subsec:sys_general}, then discuss their controllability and observability in Subsection \ref{subsec:stability_ctrb_obsv_general},  and finally study their transfer functions in Subsection \ref{subsec:tf_general}.

%%%%%%%%%%%%%%%%%%%%%%%%%%%%%%%
%%%%%%%%%%%%%%%%%%%%%%%%%%%%%%%
\subsection{Quantum linear systems} \label{subsec:sys_general}

In this subsection quantum linear systems are briefly described in terms of the 
$(S,L,H)$ language, \cite{GJ09}. More discussions on quantum linear systems can be found in, e.g., \cite{GZ00}; \cite{WM08}; \cite{WM10}; \cite{DJ99}; \cite{ZJ12}; \cite{TNP+12}.

An open quantum linear system $G$ studied in this paper consists of $n$
interacting quantum harmonic oscillators driven by $m$ input boson fields. Each oscillator $j$ has an annihilation
operator $a_{j}$ and a creation operator $a_{j}^{\ast }$; $a_{j}$
and $a_{j}^{\ast }$ are operators on the system space $\mathfrak{h}$ which
is an infinite-dimensional Hilbert space. The operators $a_{j},a_{k}^{\ast }$
satisfy the canonical commutation relations: $[a_{j},~a_{k}^{\ast }]=\delta
_{jk}$. Denote $\mathbf{a}\equiv \lbrack a_{1}~\cdots ~a_{n}]^{T}$. Then the
initial (that is, before the interaction between the system and the input boson fields) Hamiltonian $H$
can be written as $H=(1/2)\mathbf{\breve{a}}^{\dag }\Omega \mathbf{\breve{a}}
$, where $\mathbf{\breve{a}}=[\mathbf{a}^{T}~(\mathbf{a}^{\#})^{T}]^{T}$ as
introduced in the {\it Notations} part, and $\Omega =\Delta (\Omega _{-},\Omega
_{+})\in \mathbb{C}^{2n\times 2n}$ is a Hermitian matrix with $\Omega
_{-},\Omega _{+}\in \mathbb{C}^{n\times n}$. $L$ in the $(S,L,H)$ language describes the coupling of the system harmonic oscillators to the input
boson fields. The coupling is \textit{linear} and can be written as $L=[C_- \ C_+]\mathbf{\breve{a}}$ with $C_{-},C_{+}\in 
\mathbb{C}^{m\times n}$. Finally, in the linear setting $S$ in the $(S,L,H)$ language is taken to be a constant {\it unitary} matrix in $\mathbb{C}^{m\times m}$.

Each input boson field $j$ has an annihilation operator $b_{j}(t)$ and a creation operator $b_{j}^\ast(t)$, which are operators on an infinite-dimensional Hilbert space $\mathfrak{F}$. Let $\mathbf{b}(t)\equiv \lbrack b_{1}(t)~\cdots ~b_{m}(t)]^{T}$. The
operators $b_{j}(t)$ and their adjoint operators $b_{j}^{\ast }(t)$ satisfy
the following commutation relations: 
\begin{equation}
\lbrack b_{j}(t),~b_{k}^{\ast }(r)]=\delta _{jk}\delta (t-r),~~\forall
j,k=1,\ldots ,m,~\forall t,r\in \mathbb{R}.
\end{equation}
%(what is $b_j(t)$ ?????) Is $dB(t)$ always a Wiener process ?????
For each  $j=1,\ldots ,m$, the $j$-th input field can also be represented in the integral form $B_{j}(t)\equiv
\int_{0}^{t}b_{j}(r)dr$, whose Ito increment is $dB_{j}(t)\equiv   
B_{j}(t+dt)-B_{j}(t)$. Denote $B(t)\equiv \lbrack B_{1}(t)~\cdots
~B_{m}(t)]^{T}$. The gauge process can be defined by $\Lambda
_{jk}(t)=\int_{0}^{t}b_{j}^{\ast }(r)b_{k}(r)dr$, ($j,k=1,\ldots ,m$). 
% (the physical meaning of gauge process ????)
The field studied in this paper is assumed to be \textit{canonical}, that
is,  the field operators $B_{j}(t),B_{k}^{\ast
}(t),\Lambda _{rl}(t)$ satisfy the following Ito table: 
\begin{equation*}
\begin{tabular}{l|llll}
$\times $ & $dB_{k}$ & $d\Lambda _{kl}$ & $dB_{l}^{\ast }$ & $dt$ \\ \hline
$dB_{i}$ & 0 & $\delta _{ik}dB_{l}$ & $\delta _{il}dt$ & 0 \\ 
$d\Lambda _{ij}$ & 0 & $\delta _{jk}d\Lambda _{il}$ & $\delta
_{jl}dB_{i}^{\ast }$ & 0 \\ 
$dB_{j}^{\ast }$ & 0 & 0 & 0 & 0 \\ 
$dt$ & 0 & 0 & 0 & 0
\end{tabular}
\end{equation*}

Under mild assumptions, the temporal evolution of the open quantum  % what are these assumptions, Markov approximation, rotating wave approximation, and weak coupling?????
system $G$ can be described in terms of the following quantum stochastic
differential equation (QSDE): 
\begin{equation}
dU(t)=\left\{ -\left( L^{\dag }L/2+iH\right) dt+dB^{\dag }(t)L-L^{\dag
}SdB(t)+\mathrm{Tr}[(S-I)d\Lambda ^{T}(t)]\right\} U(t), ~~ t > 0,
\label{eq;U}
\end{equation}
with $U(0)=I$ being the identity operator.  Let $X$ be an operator on the system space $\mathfrak{h}$. Then the temporal
evolution of $X$, denoted $X(t)\equiv U(t)^{\ast }(X\otimes I)U(t)$, is governed by the following QSDE: 
\begin{eqnarray}
dX(t)&=&\mathcal{L}_{L,H}(X(t))dt+dB^{\dag }(t)S^{\dag
}(t)[X(t),L(t)]+[L^{\dag }(t), X(t)]S(t)dB(t) \nonumber\\
&&+\mathrm{Tr}[(S^{\dag
}(t)X(t)S(t)-X(t))d\Lambda ^{T}(t)],  \label{eq:X}
\end{eqnarray}
where the Lindblad operator $\mathcal{L}_{L,H}(X(t))$ is 
\begin{equation}
\mathcal{L}_{L,H}(X(t))\equiv -i[X(t),H(t)]+\frac{1}{2}L^{\dag
}(t)[X(t),L(t)]+\frac{1}{2}[L^{\dag }(t), X(t)]L(t).  \label{eq:L}
\end{equation}
Note that $X(t)$ is an operator on the joint system-field space $\mathfrak{h}\otimes \mathfrak{F}$.

Let $b_{out,j}(t)$ denote the $j$-th field after interacting with the
system, and $B_{out,j}(t)\equiv \int_{0}^{t}b_{out,j}(r)dr$. We have $B_{out,j}(t)=U^{\ast }(t)\left( I\otimes B_{j}(t)\right) U(t)$. Denote $B_{\mathrm{out}}(t)\equiv \lbrack B_{\mathrm{out},1}(t),~\cdots
~B_{\mathrm{out},m}(t)]^{T}$. Then in compact form the output field equation is 
\begin{equation}
dB_{\mathrm{out}}(t)=L(t)dt+SdB(t).  \label{eq:B_out}
\end{equation}
Substituting $H=(1/2)\breve{a}^{\dag }\Omega \breve{a}$ and $L=[C_- \ C_+]\mathbf{\breve{a}}$ 
into (\ref{eq:X}) we have a quantum {\it linear} system:
\begin{eqnarray}
d\mathbf{\breve{a}}(t) 
&=&
\mathcal{A}\mathbf{\breve{a}}(t)dt+\mathcal{B}d\breve{B}(t) , 
\label{eq:sys_passive_a} \\
d\check{B}_{\mathrm{out}}(t) 
&=&
\mathcal{C}\mathbf{\breve{a}}(t)dt+\mathcal{D}d\breve{B}(t) , 
\label{eq:sys_passive_b}
\end{eqnarray}
in which 
\begin{equation}
\mathcal{A}=-\frac{1}{2}C^{\flat }C-iJ_{n}\Omega ,~
\mathcal{B}=-C^{\flat }\Delta (S,0_{m\times m}),~
\mathcal{C}=\Delta(C_-,\ C_+) \equiv C,~
\mathcal{D}=\Delta (S,0_{m\times m}).
\label{eq:ABCD}
\end{equation} 
Clearly, the quantum linear system is parameterized by constant matrices 
$S,C,\Omega $. In the sequel, we use the notation $G\sim (S,C,\Omega )$ for
the quantum linear system (\ref{eq:sys_passive_a})-(\ref{eq:sys_passive_b})
with parameters given in (\ref{eq:ABCD}).

For notation's sake, we introduce the following definition.

\begin{definition}\label{def:realization_general}
 (\ref{eq:sys_passive_a})-(\ref{eq:sys_passive_b}) with parameters given in (\ref{eq:ABCD}) is said to be the {\it realization} of the quantum linear
system $G\sim (S,C,\Omega)$.
\end{definition}

The constant matrices $\mathcal{A},\mathcal{B},\mathcal{C},\mathcal{D}$ in (\ref{eq:ABCD}) satisfy the following fundamental relations:
\begin{equation}\label{eq:general:physical realizability}
\mathcal{A+A}^{\flat }+\mathcal{C}^{\flat }\mathcal{C} =0,  ~  \mathcal{B} = \mathcal{-C^{\flat }D}, ~  \mathcal{D}^{\flat }\mathcal{D} =I_{2m}.
\end{equation}
These equations are often called {\it physically realizability} conditions
of quantum linear systems. More discussions on physical realizability of quantum linear systems can be found in, e.g., \cite{JNP08}; \cite{ZJ11}; \cite{ZJ12}.

%Given a realization (\ref{eq:sys_passive_a})-(\ref{eq:sys_passive_b})  satisfying (\ref{eq:general:physical realizability}), $\Omega$ is not unique. So, there is no one-to-one correspondence between $G\sim (S,C,\Omega)$ and the realization (\ref{eq:sys_passive_a})-(\ref{eq:sys_passive_b}) if the parameters in (\ref{eq:ABCD})  are not specified.?????

%%%%%%%%%%%%%%%%%%%%%%%%%%%%%%%
%%%%%%%%%%%%%%%%%%%%%%%%%%%%%%%
\subsection{Controllability and observability} \label{subsec:stability_ctrb_obsv_general}

In this subsection we study controllability and observability of quantum linear systems introduced in Subsection \ref{subsec:sys_general}.

Let $X$ be an operator on the system space $\mathfrak{h}$. Denote by $\left\langle X(t)\right\rangle $ the
expected value of $X(t)$ with respect to the initial joint system-field state
(which is a unit vector in the Hilbert space $\mathfrak{h}\otimes \mathfrak{F}$).
Then  (\ref{eq:sys_passive_a})-(\ref{eq:sys_passive_b}) gives rise to
the following {\it classical} linear system 
\begin{eqnarray}
\frac{d\left\langle \mathbf{\breve{a}}(t)\right\rangle }{dt} 
&=&
\mathcal{A}\left\langle \mathbf{\breve{a}}(t)\right\rangle +\mathcal{B}\langle \mathbf{\breve{b}}(t)\rangle ,  \label{eq:sys_passive_average_a} \\
\frac{d\langle \mathbf{\breve{b}}_{\mathrm{out}}(t)\rangle }{dt} 
&=&
\mathcal{C}\left\langle \mathbf{\breve{a}}(t)\right\rangle +\mathcal{D}\langle \mathbf{\breve{b}}(t\rangle ).  \label{eq:sys_passive_average_b}
\end{eqnarray}

\begin{definition}\label{def:stab_ctrb_obsv} 
The quantum linear system $G\sim
(S,C,\Omega )$\ is said to be \textit{Hurwitz stable} (resp. \textit{controllable}, \textit{observable}) if the corresponding classical linear system (\ref%
{eq:sys_passive_average_a})-(\ref{eq:sys_passive_average_b}) is \textit{Hurwitz stable} (resp. \textit{controllable}, \textit{observable}).
\end{definition}

Due to the special structure of quantum linear systems, we have the following result concerning their controllability and
observability.

\begin{proposition} \label{prop:minimal_C_Omega} 
Given a quantum linear system $G\sim
(S,C,\Omega )$, the following statements are equivalent:

\begin{description}
\item[(i)] $G$ is controllable;

\item[(ii)] $G$ is observable;

\item[(iii)] $\mathrm{rank}(\mathbf{O}_s) = 2n$, where 
\begin{equation}  \label{eq:Os}
\mathbf{O}_s \equiv \left[ 
\begin{array}{c}
C \\ 
CJ_{n}\Omega \\ 
\vdots \\ 
C\left( J_{n}\Omega \right) ^{2n-1}
\end{array}
\right].
\end{equation}
\end{description}
\end{proposition}

{\bf Proof.}~ (i) $\Rightarrow $ (ii). We show this by contradiction. Assume $G$ is
not observable. By the classical control theory (see. e.g., \cite[Theorems 3.3]{ZDG96}) there exist a scalar $\lambda $ and a non-zero
vector $v\in \mathbb{C}^{2n}$ such that $\mathcal{A}v=\lambda v$ and $\mathcal{C}v=0$. So $J_{n}\Omega v=i\lambda v$ and $Cv=0$. Let $u=J_{n}v$ and $\mu=-\lambda^\ast$. Then $u^{\dagger }\mathcal{B=-}v^{\dagger }C^{\dagger }J_{m}=0$,
and 
\begin{equation*}
u^{\dagger }\mathcal{A}
=-
(J_{n}v)^{\dagger }\left( 
C^{\flat }C/2+iJ_{n}\Omega \right) 
=
\mathcal{-}(J_{n}v)^{\dagger
}iJ_{n}\Omega 
=-
iv^{\dagger }\Omega =-\lambda ^{\ast }v^{\dagger
}J_{n}=\mu u^{\dag }.
\end{equation*}
By a standard result in classical control theory, (see. e.g., \cite[Theorems 3.1]{ZDG96}), $G$ is not controllable. We reach a contradiction.

(ii) $\Rightarrow $ (i). This can be established by reversing the proof for
(i) $\Rightarrow $ (ii).

(ii) $\Rightarrow $ (iii). Let $v\in \mathbb{C}^{2n}$ such that $\mathbf{O}_{s}v=0$. Then $\mathcal{C}v=Cv=0 $ and $C(J_{n}\Omega )^{k}v=0$, $k=1,\ldots ,2n-1$.  Moreover,
\begin{eqnarray*}
\mathcal{CA}v &=&-C\left(C^{\flat }C/2+iJ_{n}\Omega
\right) v=-iCJ_{n}\Omega v=0, \\
\mathcal{CA}^{2}v &=&-C\left(C^{\flat }C/2+iJ_{n}\Omega
\right) ^{2}v=C(J_{n}\Omega )^{2}v=0, \\
&&\vdots \\
\mathcal{CA}^{2n-1}v &=&C(J_{n}\Omega )^{2n-1}v=0.
\end{eqnarray*}
But by (ii) $G$ is observable, therefore $v=0$.  (iii) is established.

(iii) $\Rightarrow $ (ii). This can be established by reversing the proof
for (ii) $\Rightarrow $ (iii).

Proposition \ref{prop:minimal_C_Omega} tells us that the controllability and
observability of a quantum linear system $G\sim (S,C,\Omega )$ are
equivalent;  moreover they can be determined by checking the rank of the matrix ${\bf O}_s$.

On the basis of Proposition \ref{prop:minimal_C_Omega}, we have the following result about the uncontrollable and unobservable subspace of a quantum linear system.

\begin{proposition}\label{prop:kernel} 
Let $\mathbf{C}\equiv [ 
\begin{array}{cccc}
\mathcal{B} & \mathcal{AB} & \cdots & \mathcal{A}^{2n-1}\mathcal{B}
\end{array}
] $ and $\mathbf{O}\equiv [ 
\begin{array}{cccc}
\mathcal{C}^{T} & \mathcal{(CA)}^{T} & \cdots & \mathcal{(\mathcal{C\mathcal{A}^{\mathrm{2n-1}}})}^{T}
\end{array}
] ^{T}$ be the controllability and observability matrices of a quantum
linear system $G\sim (S,C,\Omega )$ respectively. Then (in the terminology of modern control theory, \cite{KS72}; \cite{AV73}; \cite{Kailath80}; \cite{ZDG96})  the following
statements hold:

\begin{description}
\item[(i)] The unobservable subspace is 
\begin{equation}
\mathrm{Ker}\left( \mathbf{O}\right) = \mathrm{Ker}\left( \mathbf{O}_s\right),
\end{equation}
where $\mathrm{Ker}\left( X\right) $ denotes the null space of the matrix $X$, as introduced in the \textit{Notations} part.

\item[(ii)] The uncontrollable subspace  is 
\begin{equation}
\mathrm{Ker}\left( \mathbf{C}^{\dagger }\right) =\mathrm{Ker}\left( \mathbf{O_{s}}J_{n}\right) .  \label{eq:C_OJ}
\end{equation}

\item[(iii)] The uncontrollable and unobservable subspace is $\mathrm{Ker}\left( \mathbf{O}_{s}\right) \cap \mathrm{Ker}\left( \mathbf{O}_{s}J_{n}\right) $.
\end{description}
\end{proposition}

Proposition \ref{prop:kernel} can be established in the similar way as
Proposition \ref{prop:minimal_C_Omega}.

Propositions \ref{prop:minimal_C_Omega} and \ref{prop:kernel} appear purely
algebraic. Nevertheless, they have interesting and important physical consequences. We begin with the following lemma.

\begin{lemma}\label{lem:ctrb_obsv}
 The dimension of the space $\mathrm{Ker}\left( \mathbf{O}_{s}\right) \mathbf{\cap }\mathrm{Ker}\left( \mathbf{O}_{s}J_{n}\right)$ is even. Let the dimension of $\mathrm{Ker}\left( \mathbf{O}_{s}\right) \mathbf{\cap }\mathrm{Ker}\left( \mathbf{O}_{s}\mathbf{J}_{n}\right) $ be $2l$ for
some nonnegative integer $l$. There exists a matrix $V=[ 
\begin{array}{cc}
V_{1} & V_{2}
\end{array}
] $ with $V_{1}\in \mathbb{C}^{2n\times 2l}$ and $V_{2}\in \mathbb{C}^{2n\times 2(n-l)}$ such that
\begin{eqnarray}
\mathrm{Range}(V_{1}) 
&=&
\mathrm{Ker}\left( \mathbf{O}_{s}\right) \cap \mathrm{Ker}\left( \mathbf{O}_{s}J_{n}\right) ,  
\label{eq:W1_range} \\
 VV^{\dagger } = V^{\dagger }V 
&=&
I_{2n},  
\label{eq:W_unitary} \\
V^{\dagger }J_{n}V 
&=&
\left[ 
\begin{array}{cc}
J_{l} & 0 \\ 
0 & J_{n-l}
\end{array}
\right] .  
\label{eq:W_symplectic}
\end{eqnarray}
\end{lemma}

The proof is given in the Appendix.

We are ready to state the main result.
\begin{theorem} \label{thm:DF} 
Let $V$ be the matrix defined in Lemma \ref{lem:ctrb_obsv}. 
If $\mathrm{Range}(V_{1})$ is an invariant space under the linear
transformation of $\Omega $, then the transformed system 
\begin{equation*}
\left[ 
\begin{array}{c}
\breve{a}_{DF} \\ 
\breve{a}_{D}
\end{array}
\right] \equiv V^{\dagger }\breve{a}
\end{equation*}
has the following realization: 
\begin{eqnarray}
d\breve{a}_{DF}(t) &=&-iJ_l V_{1}^{\dagger }\Omega V_{1}\breve{a}_{DF}(t)dt,
\label{eq:DF} \\
d\breve{a}_{D}(t) 
&=&
-\left((CV_{2})^{\flat }(CV_{2})/2+iJ_{n-l} V_{2}^{\dagger }\Omega V_{2}\right) \breve{a}_{D}(t)dt-(CV_{2})^{\flat }\mathcal{D}d\check{B}(t), 
 \label{eq:D} \\
d\check{B}_{\mathrm{out}}(t) 
&=&
(CV_{2})\breve{a}_{D}(t)dt+\mathcal{D}d\check{B}(t).  
\label{eq:DF_D_output}
\end{eqnarray}
\end{theorem}

{\bf Proof.}~ Because $\mathrm{Range}(V_{1})=\mathrm{Ker}\left( \mathbf{O}_{s}\right) \cap \mathrm{Ker}\left( \mathbf{O}_{s}J_{n}\right) $, the
coupling operator of the transformed mode $[
\begin{array}{cc}
\breve{a}_{DF}^{T} & \breve{a}_{D}^{T}
\end{array}
]^{T}$ is $\mathcal{C}V=[ 
\begin{array}{cc}
0 &CV_{2}
\end{array}
] $.
Moreover, because $\mathrm{Range}(V_{1})$ is an invariant space under the
linear transformation of $\Omega $, there exists a matrix $Y$ such that $\Omega  V_1 = V_1 Y$. We have $V_1^\dag  \Omega V_2 = Y^\dag V_1^\dag V_2 =0$ where (\ref{eq:W_unitary}) is used. This, together with (\ref{eq:W_symplectic}), gives   
\begin{equation*}
V^{\dagger }J_n\Omega V=V^{\dagger }J_n V V^\dag \Omega V =\left[ 
\begin{array}{cc}
J_l V_{1}^{\dagger }\Omega V_{1} & 0 \\ 
0 & J_{n-l} V_{2}^{\dagger }\Omega V_{2}
\end{array}
\right] .
\end{equation*}
That is, the transformed system with mode $[
\begin{array}{cc}
\breve{a}_{DF}^{T} & \breve{a}_{D}^{T}
\end{array}
]^{T}$ has the realization (\ref{eq:DF})-(\ref{eq:DF_D_output}).

{\it Remark 1.}  By (\ref{eq:DF}), the modes $\breve{a}_{DF}$ evolve unitarily as an isolated system. In literature such isolated modes embedded in an open quantum system is often called {\it decoherence-free} modes, see, e.g.,  \cite{TV08}, \cite{TV09}, \cite{NY13}. Theorem \ref{thm:DF}  can be viewed as the complex-domain counterpart of Theorem 3.1 in \cite{NY13} in the real domain. However, with the help of the matrix ${\bf O}_s$, matters are simplified; moreover, it can be seen from the proof of Lemma \ref{lem:ctrb_obsv} in the Appendix that the structure of the unitary transformation matrix $V$ is better revealed with the help of ${\bf O}_s$ and in the complex domain.

Finally, from the proof of Lemma \ref{lem:ctrb_obsv} it can be seen that the dimension of the space ${\rm Ker}(C)$ is also even. Moreover we have the following corollary which shows that under some conditions the unobservable and uncontrollable subspace is exactly ${\rm Ker}(C)$.
\begin{corollary}\label{cor:DF}
Let the dimension of the space ${\rm Ker}(C)$ be $2r$. Let a matrix $T\in \mathbb{C}^{2n\times 2r}$ be such that ${\rm Range}(T)={\rm Ker}(C)$. If $J_n T = T J_r$ and   ${\rm Range}(T)$ is an invariant space under the linear transformation of $\Omega$, then ${\rm Ker}(C) = {\rm Ker}({\bf O_s})\cap {\rm Ker}({\bf O_s}J_n)$.
\end{corollary}

{\bf Proof.}~ Clearly, $\mathrm{Ker}\left( \mathbf{O}_{s}\right) \mathbf{\cap }\mathrm{Ker}\left( \mathbf{O}_{s}J_{n}\right) \subset  {\rm Ker}({\bf O}_s) \subset \mathrm{Ker}\left(C\right) $. Thus it is sufficient to show that $\mathrm{Ker}\left( C\right)
\subset \mathrm{Ker}\left( \mathbf{O}_{s}\right) \mathbf{\cap }\mathrm{Ker}\left( \mathbf{O}_{s}J_{n}\right) $. However $\mathrm{Range}(T)=\mathrm{Ker}\left( C\right) $, we show that $\mathrm{Range}(T)\subset 
\mathrm{Ker}\left( \mathbf{O}_{s}\right) \mathbf{\cap }\mathrm{Ker}\left( 
\mathbf{O}_{s}J_{n}\right) $. Because $\mathrm{Range}(T)$ is invariant
with respect to a linear transformation $\Omega$, there exist matrix $Y$
such that $\Omega T=TY$. This, together with $J_{n}T=TJ_{r}$, gives  $C(J_{n}\Omega )T = CTJ_r Y=0$.   
Similarly, for all $k\geq 1$,  $\mathcal{C}(J_{n}\Omega )^{k}T=0$. That is, ${\bf O}_s T=0$. Moreover, ${\bf O}_s J_n T={\bf O}_s T J_r=0$. Consequently $ {\rm Ker}(C)= \mathrm{Range}(T)\subset \mathrm{Ker}\left( \mathbf{O}_{s}\right) \mathbf{\cap }\mathrm{Ker}\left( \mathbf{O}_{s}J_{n}\right) $. This together with $\mathrm{Ker}\left( \mathbf{O}_{s}\right) \mathbf{\cap }\mathrm{Ker}\left( \mathbf{O}_{s}J_{n}\right)\subset \mathrm{Ker}\left( C\right) $ yields $\mathrm{Ker}\left( C\right) =\mathrm{Ker}\left( \mathbf{O}_{s}\right) \mathbf{\cap }\mathrm{Ker}\left( 
\mathbf{O}_{s}J_{n}\right) $. 

Corollary \ref{cor:DF} can be regarded as the complex-domain counterpart of Proposition 3.1 in \cite{NY13} in the real domain.

%%%%%%%%%%%%%%%%%%%%%%%%%%%%%%%
%%%%%%%%%%%%%%%%%%%%%%%%%%%%%%%
\subsection{Transfer functions} \label{subsec:tf_general}

In the frequency domain, the transfer function  of
the system $G\sim (S,C,\Omega )$ is defined to be 
\begin{equation}  \label{tf_general}
G(s)\equiv \mathcal{D}+\mathcal{C}(sI-\mathcal{A})^{-1}\mathcal{B}.
\end{equation}
This transfer function has the following fundamental property, see, e.g., \cite[Eq. (24)]{ZJ13}:
\begin{equation}  \label{eq:mar12_1}
G(i\omega)^\flat G(i\omega) = G(i\omega) G(i\omega)^\flat= I_{2m}, \ \ \
\forall \omega\in\mathbb{R}.
\end{equation}
Interestingly, the transfer function $G(s)$ of the quantum linear system $G\sim (S,C,\Omega )$ can be written into a fractional form.

\begin{proposition}\label{lem:G_Sigma_0}
 The transfer function $G(s)$ for a Hurwitz stable quantum linear
system $G\sim (S,C,\Omega )$ can be written in the following fractional  form 
\begin{equation}
G(s)=(I-\Sigma (s))(I+\Sigma (s))^{-1} \Delta(S,0),  \label{eq:G_general}
\end{equation}
where 
\begin{equation}
\Sigma (s)\equiv \dfrac{1}{2}C(sI+iJ_n\Omega )^{-1}C^{\flat },~~\forall 
\mathrm{Re}[s]>0.  \label{eq:G_Sigma}
\end{equation}
\end{proposition}

{\bf Proof.}~ 
Because the system $G(s)$ is Hurwitz stable, all the eigenvalues of the
matrix $A$ have strictly negative real part, therefore the matrix $sI-A$ is
invertible for all \textrm{Re}$[s]>0$. Moreover, for all \textrm{Re}$[s]>0$,
by the Woodbury matrix inversion formula,
\begin{eqnarray*}
(sI-\mathcal{A})^{-1} &=&(sI+iJ_{n}\Omega +\frac{1}{2}C^{\flat }C)^{-1} \\
&=&\left( sI+iJ_{n}\Omega \right) ^{-1}-\frac{1}{2}\left( sI+iJ_{n}\Omega
\right) ^{-1}C^{\flat }\left( I+\frac{1}{2}C\left( sI+iJ_{n}\Omega \right)
^{-1}C^{\flat }\right)^{-1} C\left( sI+iJ_{n}\Omega \right) ^{-1}.
\end{eqnarray*}
As a result, for all \textrm{Re}$[s]>0$,
\begin{eqnarray*}
&&I-C(sI-\mathcal{A})^{-1}C^{\flat } \\
&=&I-C\left\{ \left( sI+iJ_{n}\Omega \right) ^{-1}-\frac{1}{2}\left(
sI+iJ_{n}\Omega \right) ^{-1}C^{\flat }(I+\frac{1}{2}C\left( sI+iJ_{n}\Omega
\right) ^{-1}C^{\flat })^{-1}C\left( sI+iJ_{n}\Omega \right) ^{-1}\right\}
C^{\flat } \\
&=&I-2\Sigma (s)+2\Sigma (s)\left( I+\Sigma (s)\right) ^{-1}\Sigma (s) \\
&=&(I-\Sigma (s))(I+\Sigma (s))^{-1},
\end{eqnarray*}
with $\Sigma (s)$ as defined in (\ref{eq:G_Sigma}). Consequently, 
\[
G(s)= (I-C(sI-\mathcal{A})^{-1}C^{\flat })\Delta(S,0) = (I-\Sigma (s))(I+\Sigma (s))^{-1} \Delta(S,0).
\]

%\begin{corollary}
%\label{cor:Sigma}If $\Omega _{+}=0$, then $\Sigma (i\omega )+\Sigma (i\omega
%)^{\flat }=0$, where  $i\omega $ is not a pole of $\Sigma (s)$ defined in (%
%\ref{eq:G_Sigma}).
%\end{corollary}
%
%Proof. Because 
%\begin{eqnarray}
%\Sigma (s)+\Sigma (s)^{\flat } &=&\dfrac{1}{2}C(sI+i\Omega )^{-1}C^{\flat }+%
%\dfrac{1}{2}C(s^{\ast }I-iJ\Omega J)^{-1}C^{\flat }  \nonumber \\
%&=&\dfrac{1}{2}C(sI+i\Omega )^{-1}\left\{ (s+s^{\ast })I+2i\Delta \left(
%0,\Omega _{+}\right) \right\} \left( C(sI+i\Omega )^{-1}\right) ^{\flat }.
%\label{eq:temp2}
%\end{eqnarray}%
%Assume that $i\omega $ is not a pole of $\Sigma (s)$. Substituting $i\omega $
%into (\ref{eq:temp2}) yields 
%\begin{equation*}
%\Sigma (i\omega )+\Sigma (i\omega )^{\flat }=2iC(sI+i\Omega )^{-1}\Delta
%\left( 0,\Omega _{+}\right) \left( C(sI+i\Omega )^{-1}\right) ^{\flat }.
%\end{equation*}%
%If $\Omega _{+}=0$, then%
%\begin{equation*}
%\Sigma (i\omega )+\Sigma (i\omega )^{\flat }=0.
%\end{equation*}
%
%\begin{remark}
%According to Corollary \ref{cor:Sigma}, that $\Sigma (i\omega )+\Sigma
%(i\omega )^{\flat }=0$ does not mean that both $\Omega _{+}=0$ and $C_{+}=0$.
%\end{remark}

%%%%%%%%%%%%%%%%%%%%%%%%%%%%
%%%%%%%%%%%%%%%%%%%%%%%%%%%%
%%%%%%%%%%%%%%%%%%%%%%%%%%%%
\section{Quantum linear passive systems}\label{sec:passive}

In this section quantum linear passive systems are studied. This type of systems is introduced in Subsection \ref{subsec:passive}. Stability, controllability and observability are investigated in Subsection \ref{sec:stability_minimality}, while minimal realizations of quantum linear passive systems are studied in Subsection \ref{subsec:minimal}. The relation between $G$ and $\Sigma$ in the passive setting is discussed in Subsection \ref{sec:dissipation}.

%%%%%%%%%%%%%%%%%%%%%%%%%%%%
%%%%%%%%%%%%%%%%%%%%%%%%%%%%
\subsection{Quantum linear passive systems} \label{subsec:passive}

If the matrices $C_{+}=0$ and $\Omega _{+}=0$, the resulting system,
parameterized by matrices $S,~C_{-,}~\Omega _{-}$, is often said to be a
quantum linear \textit{passive} system. In this case, it can be
described entirely in terms of annihilation operators. Actually a quantum
linear passive system has the following form: 
\begin{eqnarray}
d\mathbf{a}(t) &=&A\mathbf{a}(t)-C_{-}^{\dag }SdB(t),  \label{eq:AB} \\
dB_{out}(t) &=&C_{-}\mathbf{a}(t)+SdB(t).  \label{eq:CD}
\end{eqnarray}
in which $A\equiv -\frac{1}{2}C_{-}^{\dag }C_{-}-i\Omega _{-}$.

In analog to Definition \ref{def:realization_general} for realization of
general linear systems we introduce the following realization concept for
passive linear systems.

\begin{definition}
\label{def:realization_passive} (\ref{eq:AB})-(\ref{eq:CD}) is said to be
the {\it realization} of the quantum linear passive system $G\sim (S,C_{-},\Omega _{-})$.
\end{definition}

Clearly, the transfer function of $G\sim (S,C_{-},\Omega _{-})$ is
\begin{equation}
G(s)=S-C_{-}(sI-A)^{-1}C_{-}^{\dag }S.  \label{tf_passive}
\end{equation}
Define 
\begin{equation}\label{sigma_passive}
\Sigma (s)\equiv \dfrac{1}{2}C_{-}(sI+i\Omega _{-})^{-1}C_{-}^{\dagger } .
\end{equation}
Then, in analog to Proposition \ref{lem:G_Sigma_0},  we have 
\begin{equation}
G(s)=(I-\Sigma (s))(I+\Sigma (s))^{-1}S .  \label{eq:G_sigma}
\end{equation}
In the passive case, Eq. (\ref{eq:mar12_1}) reduces to 
\begin{equation}  \label{eq:mar12_2}
G(i\omega)^\dag G(i\omega) = G(i\omega) G(i\omega)^\dag= I_m, \ \ \ \forall
\omega\in\mathbb{R}.
\end{equation}

Because deferent realizations may correspond to the same transfer
function (\ref{tf_passive}),  we introduce the following concept.

\begin{definition}
\label{def:unitary} Two realizations are said to be \textit{unitarily
equivalent} if there exists a unitary transformation which transforms one to
the other.
\end{definition}

Clearly, two unitarily equivalent realizations correspond to the same transfer
function.

%%%%%%%%%%%%%%%%%%%%%%%%%%%%%%%%
%%%%%%%%%%%%%%%%%%%%%%%%%%%%%%%%
\subsection{Stability, controllability, and observability} \label{sec:stability_minimality}

In this subsection we study stability of quantum linear passive systems. In
particular, we show that a quantum linear passive
system $G\sim (S,C_{-},\Omega _{-})$ is Hurwitz stable if and only if it is observable and controllable. 

\begin{lemma}\label{thm:stab_obsr_cont}
 The following statements for a quantum linear
passive system $G\sim (S,C_{-},\Omega _{-})$ are equivalent:

\begin{description}
\item[(i)] $G$ is Hurwitz stable;

\item[(ii)] $G$ is observable;

\item[(iii)] $G$ is controllable.
\end{description}
\end{lemma}

{\bf Proof.}~ 
(i) $\rightarrow $ (ii). Clearly, $X=I_{n}>0$ is the
unique solution to the following Lyapunov equation 
\begin{equation}
A^{\dag }X+XA+C_{-}^{\dag }C_{-}=0.  \label{eq:Lyap}
\end{equation}
According to \cite[Lemma 3.18]{ZDG96}, $(C_{-}^{\dag }C_{-},A)$ is
observable, so $(C_{-},A)$ is observable. That is, $G$ is observable.

(ii) $\rightarrow $ (i). Because $X=I_{n}>0$ is a solution
to Eq. (\ref{eq:Lyap}), $C_{-}^{\dag }C_{-}\geq 0$ and $(C_{-}^{\dag
}C_{-},A)$ is observable, by \cite[Lemma 3.19]{ZDG96}, $A$ is Hurwitz stable.

The equivalence between (ii) and (iii) has been established in Proposition \ref{prop:minimal_C_Omega}.

{\it Remark 2.}  An alternative proof of the equivalence between (ii) and (iii) is given in \cite[Lemma 3.1]{GY13}. An alternative proof of (ii) $\rightarrow $ (i) is given in \cite[Lemma 3.2]{GY13}.

%%%%%%%%%%%%%%%%%%%%%%%%%%%%%%%%%%%
%%%%%%%%%%%%%%%%%%%%%%%%%%%%%%%%%%%
\subsection{Minimal realization} \label{subsec:minimal}

In this subsection we study minimal realization of a given quantum linear
passive system $G\sim (S,C_{-},\Omega _{-})$. We first introduce the concept
of minimal realization.

\begin{definition}
If a quantum linear passive system $G\sim (S,C_{-},\Omega _{-})$ is both 
controllable and observable, we say its realization (\ref{eq:AB})-(\ref{eq:CD}) is a \textit{minimal} realization.
\end{definition}

The following result is an immediate consequence of Lemma \ref{thm:stab_obsr_cont}.

\begin{theorem}\label{thm:minimal_hurwitz}
 (\ref{eq:AB})-(\ref{eq:CD}) is a minimal
realization of the quantum linear passive system $G\sim (S,C_{-},\Omega
_{-}) $ if and only if it is Hurwitz stable.
\end{theorem}

In what follows we study the following problem concerning minimal realization.

\begin{problem}
Given a quantum linear passive system  $G\sim (S,C_{-},\Omega_{-})$ which may not be Hurwitz stable, it may have a subsystem  $ (S,C_{\mathrm{min}},\Omega_{\mathrm{min}})$ which is Hurwitz stable. In this case, let $n_{\mathrm{min}}$ be the number of system oscillators in the minimal
realization of $ (S,C_{\mathrm{min}},\Omega_{\mathrm{min}})$. How to
compute $n_{\mathrm{min}}$?
\end{problem}

%%%%%%%%%%%%%%%%%%%%%%%%%%%%%%%%%
\subsubsection{The single-input-single-output (SISO) case}

Given a SISO quantum linear passive system $G(s)$, let  the
spectral decomposition of $\Omega_- $ be 
\begin{equation*}
\Omega_- =\sum_{\omega \in \mathrm{spec}(\Omega_- )}\omega P_{\omega },
\end{equation*}
where $P_{\omega }$ denotes the projection onto the eigenspace of the
eigenvalue $\omega $ of $\Omega _{-}$. Define 
\begin{equation}
\sigma (\Omega _{-},C_{-})\equiv \{\omega \in \mathrm{spec}(\Omega
_{-}):C_{-}P_{\omega }C_{-}^{\dag }\neq 0\}.  \label{eq:spectral}
\end{equation}

The following result shows that the size of the set $\sigma (\Omega
_{-},C_{-})$ is nothing but $n_{\mathrm{min}}$.

\begin{proposition}\label{prop:siso_minimal}
 Given a SISO quantum linear passive system $G\sim
(S,C_{-},\Omega _{-})$, the number $n_{\text{min}}$ of oscillators of a
minimal realization $(S,C_{\text{min}},\Omega _{\text{min}})$ is equal to
the size of the set $\sigma (\Omega _{-},C_{-})$ defined in (\ref{eq:spectral}).
\end{proposition}

The proof is given in the Appendix.

%%%%%%%%%%%%%%%%%%%%%%%%%
\subsubsection{The multi-input-multi-output (MIMO) case}
The following result is the MIMO version of Proposition \ref{prop:siso_minimal}.
\begin{proposition}\label{prop:mimo_minimal}
 For a MIMO quantum linear passive system $G\sim
(S,C_{-},\Omega _{-})$, let the distinctive eigenvalues of $\Omega _{-}$ be 
$\omega _{1},\ldots ,\omega _{r}$, each with algebraic multiplicity $\tau
_{i} $ respectively, $i=1,\ldots ,r$. Define $\Lambda _{i}=\omega
_{i}I_{\tau _{i}}$, $i=1,\ldots ,r$. Assume 
\begin{equation}
\Omega _{-}=\left[ 
\begin{array}{ccc}
\Lambda _{1} &  & 0 \\ 
& \ddots &  \\ 
0 &  & \Lambda _{r}
\end{array}
\right].
\end{equation}
Accordingly partition $C_{-}=[C_{1}~~C_{2}~~\cdots ~~C_{r}]$ with $C_{i}$
having $\tau _{i}$ columns, $i=1,\ldots ,r$. Then 
\begin{equation*}
n_{\mathrm{min}}=\sum_{i=1}^{r}\mathrm{column~rank}[C_{i}]
\end{equation*}
In particular, if $\tau _{i}=1$ for all $i=1,\ldots ,r$, that is, all poles
of $\Omega _{-}$ are simple poles, then 
\begin{equation}
n_{\mathrm{min}}=\{\omega _{i}\in \mathrm{spec}(\Omega _{-}):\mathrm{Tr}
[C_{-}P_{\omega _{i}}C_{-}^{\dag }]\neq 0\},  \label{eq:spectrm_C_mimo}
\end{equation}
as given in Proposition \ref{prop:siso_minimal}.
\end{proposition}

The construction in Proposition \ref{prop:mimo_minimal} is essentially the
Gilbert's realization. Its proof follows the discussions in \cite[Sec. 6.1]{Kailath80} or \cite[Sec. 3.7]{ZDG96}. The details are omitted.

%%%%%%%%%%%%%%%%%%%%%%%%%%%%%%%%%%%%%
%%%%%%%%%%%%%%%%%%%%%%%%%%%%%%%%%%%%%
\subsection{$G$ and $\Sigma$}\label{sec:dissipation}
In this subsection we explore a further relation between a quantum linear passive system $G\sim(S,C_-,\Omega_-)$ and $\Sigma$ defined in (\ref{sigma_passive}). 

We first review the notions of lossless bounded real and lossless positive real.  The bounded real lemma for quantum linear passive systems has been established in \cite{MP11a}. Dissipation theory for more general quantum linear systems has been studied in \cite{JNP08}, \cite{ZJ11}, while the nonlinear case has been studied in \cite{JG10}.

\begin{definition} \label{def:lossless_bounded} (Lossless Bounded Real, \cite[Definition 6.3]{MP11a}.) 
A quantum linear passive system $G=(S,C_{-},\Omega _{-})$ is said to be lossless bounded real if it is Hurwitz stable and Eq. (\ref{eq:mar12_2}) holds. 
\end{definition}

According to Definition \ref{def:lossless_bounded}, a Hurwitz stable quantum
linear passive system is naturally lossless bounded real, as derived in \cite{MP11a}.  

Positive real functions have been studied extensively in classical (namely, non-quantum) control theory, see, e.g., \cite{AV73}. Here we state a complex-domain version of positive real functions.

\begin{definition} \label{def:positive}(Lossless Positive Real.) 
A function $\Xi (s)$ is said to be
positive real if it is analytic in $\mathrm{Re}[s]>0$ and satisfies 
\begin{equation*}
\Xi (s)+\Xi (s)^{\dag }\geq 0,~~~\forall \mathrm{Re}[s]>0.
\end{equation*}
Moreover, $\Xi(s)$ is called {lossless positive real} if is positive real and
satisfies 
\begin{equation}
\Xi (i\omega )+\Xi (i\omega )^{\dag }=0,
\end{equation}
where $i\omega $ is not a pole of $\Xi(s)$.
\end{definition}

The following result relates the lossless bounded realness of a quantum
linear passive system $G\sim (S,C_{-},\Omega _{-})$ to the lossless positive
realness of $\Sigma (s)$ defined in Eq. (\ref{sigma_passive}).

\begin{theorem}\label{eq:G_Sigma_Z} 
If a quantum linear passive system $G\sim(S,C_{-},\Omega _{-})$ is minimal, then

\begin{description}
\item[(i)] $G(s)$ is lossless bounded real.

\item[(ii)] $\Sigma (s)$ defined in Eq. (\ref{sigma_passive}) is lossless
positive real.
\end{description}
\end{theorem}

{\bf Proof.}~ 
(i). Without loss of generality, assume $S=I_m$. Because $G\sim
(I,C_{-},\Omega _{-})$ is minimal, by Theorem \ref{thm:minimal_hurwitz}, it
is Hurwitz stable. Moreover, $G\sim (I,C_{-},\Omega _{-})$ satisfies Eq. (\ref{eq:mar12_2}). Therefore, according to Definition \ref{def:lossless_bounded}, $G\sim (I,C_{-},\Omega _{-})$ is lossless bonded
real.

(ii). Assume $i\omega $ is not a pole of $\Sigma (s)$. Then the matrix $i\omega
I+i\Omega _{-}$ is invertible. Note that
\begin{eqnarray}
\Sigma (s)+\Sigma (s)^{\dag } &=&\dfrac{1}{2}C_{-}(sI+i\Omega
_{-})^{-1}C_{-}^{\dagger }+\dfrac{1}{2}C_{-}(s^{\ast }I-i\Omega
_{-})^{-1}C_{-}^{\dagger }  \label{eq:temp1} \\
&=&\text{Re}\left[ s\right] C_{-}(sI+i\Omega _{-})^{-1}\left(
C_{-}(sI+i\Omega _{-})^{-1}\right) ^{\dagger },~~~\forall \mathrm{Re}[s]>0. 
\notag
\end{eqnarray}
 By (\ref{eq:temp1}), $\Sigma (i\omega
)+\Sigma (i\omega )^{\dag }=0$. Therefore, by Definition \ref{def:positive}, 
$\Sigma (s)$ is lossless positive real.

{\it Remark 3.} 
In fact it can be shown that in the minimal realization case (i) and (ii) in Theorem \ref{eq:G_Sigma_Z} are equivalent.

{\it Remark 4.} 
$\Sigma (s)$ is a linear port-Hamiltonian system, \cite[Chapter 4]{vdS96}.
However it is worth noting that a lossless positive real $\Sigma (s)$ may
not generate a genuine quantum system $G(s)$ via $G(s)=(I-\Sigma
(s))(I+\Sigma (s))^{-1}$. For example, given $\Sigma (s)=\frac{s^{2}+1}{s(s^{2}+2)}$, it is lossless positive real. However, it can be verified that 
$G(s)=(I-\Sigma (s))(I+\Sigma (s))^{-1}=1-\frac{2(s^{2}+1)}{s^{3}+s^{2}+2s+1}
$ is not a genuine quantum linear system. Later in Section \ref{sec:equivalence} we will give an explicit form of $\Sigma (s)$ which
generates a genuine quantum linear passive system $G$, see (\ref{eq:Sigma_Delta}) for details.

Here we have used the annihilation-operator form to study dissipative
properties of quantum linear passive systems. Because the resulting matrices may be
complex-valued, they can be viewed as the complex versions of lossless
bounded real and lossless positive real in terms of the quadrature
form, \cite{JNP08}. In fact, if the quantum system is represented in the
quadrature form, it is exactly the same lossless bounded real form as that
in \cite[Secs. 2.6 and 2.7]{AV73} for classical linear systems. In fact, the relation between lossless bounded real and lossless positive real is well-known in electric networks, see. e.g., \cite{AV73}.

\section{Realizations for quantum linear passive systems} \label{sec:representation}

Several realizations of quantum linear passive systems are proposed in
this section. The multi-input-multi-output (MIMO) case is studied in
Subsection \ref{subset:MIMO}. For the single-input-single-output (SISO)
case, an independent-oscillator realization is proposed in Subsection \ref{sec:IO}, Fig.~\ref{fig:rep1}; a chain-mode realization is presented in Subsections \ref{subset:chain}, Fig.~\ref{fig:repCM}; and the uniqueness of the independent-oscillator realization is discussed in Subsection \ref{sec:equivalence}.

%%%%%%%%%%%%%%%%%%%%%%%%%%%%%%%%
%%%%%%%%%%%%%%%%%%%%%%%%%%%%%%%%
\subsection{Realizations for multi-input-multi-output models} \label{subset:MIMO}

In this subsection a new realization for MIMO quantum linear passive systems is proposed.

Before presenting our realizations for quantum linear passive systems, we
describe for completeness a realization proposed in \cite{HIN10} and \cite%
{IRP11} using the series product to produce a realization of an $n$-oscillator 
system as a cascade of $n$ one-oscillator systems.

\begin{figure}[htbp]
\centering
\includegraphics[width=0.50\textwidth]{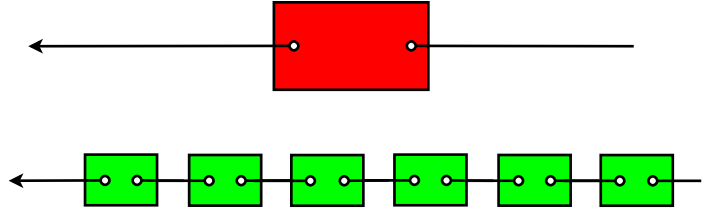} 
\label{fig:repcasc}
\caption{A quantum linear passive system with $n$ system oscillators is realised  as a sequence of $n$ components in series, each one having a one-mode oscillator.}
\end{figure}

We begin with the observation that every matrix $n\times n$ matrix $A$
admits a Schur decomposition $A=U^{\dagger }A^{\prime }U$ with $U$ unitary and $A^{\prime }$ lower triangular. For a given quantum linear passive system $G\sim (S,C_{-},\Omega _{-})$, we define a unitary
transform $a^{\prime }\equiv Ua$, such that $A^{\prime }=UAU^\dag$ is lower
triangular. Accordingly denote $C^{\prime }=C_{-}\,U^{\dagger }$ and $\Omega
^{\prime }=U\Omega _{-}U^{\dagger }$. The new system is thus $G^{\prime
}\sim (S,C^{\prime },\Omega ^{\prime })$. A standard result from linear
systems theory shows that the two systems $G$ and $G^{\prime }$ have the
same transfer function. In what follows we show the system $G^{\prime }$ has a cascade
realization, Fig. \ref{fig:repcasc}. Because $A^{\prime }=-\frac{1}{2}C^{\prime \dagger
}C^{\prime }-i\Omega ^{\prime }$ is lower triangular, for $j<k$ we have $%
A_{jk}^{\prime }=-\frac{1}{2}C_{j}^{\prime \dagger }C_{k}^{\prime }-i\Omega
_{jk}^{\prime }=0$, so $\Omega _{jk}^{\prime }=\frac{i}{2}C_{j}^{\prime
\dagger }C_{k}^{\prime }$. Therefore the lower triangular components are 
\begin{equation*}
A_{kj}^{\prime }=-\frac{1}{2}C_{k}^{\prime \dagger }C_{j}^{\prime }-i\Omega
_{kj}^{\prime }=-\frac{1}{2}C_{k}^{\prime \dagger }C_{j}^{\prime }-i\Omega
_{jk}^{\prime \ast }\equiv -C_{k}^{\prime \dagger }C_{j}^{\prime }.
\end{equation*}

Let us now set $G_{0}\sim (S,0,0)$ and $G_{k}\sim (I,C_{k}^{\prime },\Omega
_{kk}^{\prime })$ then the new system $G^{\prime }$ has a the cascaded
realization $G^{\prime }=G_{n}\vartriangleleft \cdots \vartriangleleft
G_{1}\vartriangleleft G_{0}$, Fig. \ref{fig:repcasc}.

Next we present a new realization for MIMO quantum linear passive systems, which may have: 1) a set of
inter-connected principal oscillators $\tilde{a}_{\mathrm{pr}}$ that interact
with the (possibly part of) environment $\tilde{b}_{\mathrm{pr}}(t)$; 2)
auxiliary oscillators $\tilde{a}_{\mathrm{aux,1}}$ and $\tilde{a}_{\mathrm{aux,2}}$) which only couple to the principal oscillators while otherwise
being independent; 3) input-out channels $\tilde{b}_{\mathrm{aux}}(t)$ that
do not couple to the system oscillators.

\begin{theorem}\label{thm:MIMO}
 A quantum linear passive system $G=(I,C_{-},\Omega _{-})$
can be unitarily transformed to another one with the corresponding realization 
\begin{eqnarray}
d\tilde{a}_{\mathrm{pr}}(t) 
&=&
-(\frac{\sigma (C_{-})^{2}}{2}+i\tilde{\Omega}_{1})\tilde{a}_{\mathrm{pr}}(t)dt-i\tilde{\Omega}_{21}\tilde{a}_{\mathrm{aux},1}(t)dt-i\tilde{\Omega}_{22}\tilde{a}_{\mathrm{aux},2}(t)dt-\sigma
(C_{-})d\tilde{B}_{\mathrm{in,pr}}(t),  
\label{eq:system3_a} \\
d\tilde{a}_{\mathrm{aux},1}(t) 
&=&
-i\sigma (\tilde{\Omega}_{3})\tilde{a}
_{\mathrm{aux},1}(t)dt-i\tilde{\Omega}_{21}^{\dag }\tilde{a}_{\mathrm{pr}}(t)dt,
\label{eq:system3_b} \\
d\tilde{a}_{\mathrm{aux},2}(t) 
&=&
-i\tilde{\Omega}_{22}^{\dag }\tilde{a}_{\mathrm{pr}}(t)dt, 
 \label{eq:system3_c} \\
dB_{\mathrm{out,pr}}(t) 
&=&
\sigma (C_{-})\tilde{a}_{\mathrm{pr}}(t)dt+dB_{\mathrm{in,pr}}(t),  
\label{eq:system3_d} \\
dB_{\mathrm{out,aux}}(t) 
&=&
dB_{\mathrm{in,aux}}(t),
\label{eq:system3_e}
\end{eqnarray}
where $\tilde{\Omega}_{1}=\tilde{\Omega}_{1}^{\dag }$, $\tilde{\Omega}_{3}=\tilde{\Omega}_{3}^{\dag }$, and $\sigma (X)$ denotes the diagonal matrix
with diagonal entries being the non-zero singular values of the matrix $X$.
Clearly, this new realization corresponds the a quantum linear passive system $\left( I,\bar{C},\bar{\Omega}\right) $ with
\begin{equation}
\bar{C} \equiv \left[ 
\begin{array}{ccc}
\sigma (C_{-}) & 0 & 0 \\ 
0 & 0 & 0
\end{array}
\right] , \ \ \ \bar{\Omega} \equiv \left[ 
\begin{array}{ccc}
\tilde{\Omega}_{1} & \tilde{\Omega}_{21} & \tilde{\Omega}_{22} \\ 
\tilde{\Omega}_{21}^{\dag } & \sigma (\tilde{\Omega}_{3}) & 0 \\ 
\tilde{\Omega}_{22}^{\dag } & 0 & 0
\end{array}
\right] .  \label{eq:C_Omega}
\end{equation}
\end{theorem}

The proof is given in the Appendix.

The realization (\ref{eq:system3_a})-(\ref{eq:system3_e}) is in some sense like controllability and observability decomposition of quantum linear passive systems. In fact by Proposition \ref{prop:minimal_C_Omega}  and Theorem \ref{thm:DF}, we have the following result.

\begin{corollary}
For the realization (\ref{eq:system3_a})-(\ref{eq:system3_e}),

\begin{enumerate}
\item the mode $\tilde{a}_{\mathrm{pr}}$ is both controllable and observable;

\item if the system $G=(I,C_{-},\Omega _{-})$ is Hurwitz stable, then $\tilde{\Omega}_{21}\neq 0$ and $\tilde{\Omega}_{22}\neq 0$.
\end{enumerate}
\end{corollary}

{\it Remark 5.}  When $m=1$, assuming minimal realization, from the
proof given in the Appendix it can be seen that Theorem \ref{thm:MIMO}
reduces to Theorem \ref{prop:independent_oscillator}  for the independent-oscillator
realization of SISO systems to be discussed in Subsection \ref{sec:IO}.

%%%%%%%%%%%%%%%%%%%%%%%%%%%%%%%%
%%%%%%%%%%%%%%%%%%%%%%%%%%%%%%%%
\subsection{Realizations for single-input-single-output models}
In this subsection, two realizations, namely the independent-oscillator realization and the chain-mode realization, of SISO quantum linear passive systems are proposed.

%%%%%%%%%%%%%%%%%%%%%%%%%%%%%%%%
\subsubsection{Independent-oscillator realization}\label{sec:IO}

\begin{figure}[tbph]
\centering
\includegraphics[width=0.500\textwidth]{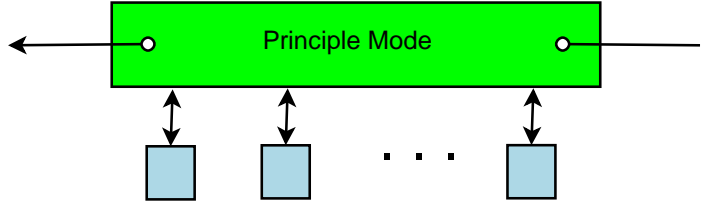}
\caption{The independent-oscillator realization: the principal mode is
coupled to $n-1$ independent auxiliary modes. The principal mode couples to
the field, while the auxiliary modes are independent other than that they
couple to the principal mode.}
\label{fig:rep1}
\end{figure}

Given a SISO quantum linear passive system $G\sim (I,C_{-},\Omega _{-})$
where 
\begin{equation} \label{eq:independent_original_osillator_parameters}
C_{-}=[\sqrt{\gamma _{1}}\ \ldots \ \sqrt{\gamma _{n}}],~~\Omega _{-}=(\omega
_{jk})_{n\times n},  
\end{equation}
 we show how to find a unitarily equivalent realization in terms of a single oscillator
(the coupling mode $c_{0} $, we also call it the principle mode) which is
then coupled to $n-1$ auxiliary modes $c_{1},\cdots ,c_{n-1}$. The auxiliary
modes are themselves otherwise independent oscillators, Fig.~\ref{fig:rep1}.

\begin{theorem}\label{prop:independent_oscillator} 
There exists a unitary matrix $T$ such
that the transformed modes 
\begin{equation} \label{eq:c}
\mathbf{c} = \left[ 
\begin{array}{c}
c_{0} \\ 
c_{1} \\ 
\vdots \\ 
c_{n-1}
\end{array}%
\right] \equiv T\,\mathbf{a}
\end{equation}
have the following realizations
\begin{eqnarray}
dc_0(t) 
&=&
-(\gamma/2+i\omega_0)c_0(t)dt -\sum_{j=1}^{n-1} i\sqrt{\kappa_j}c_j(t) -\sqrt{\gamma}dB(t),
 \label{eq:L_independent_oscillator_L}  \\ 
dc_j(t)
&=&
-i\omega_j c_j(t)dt-i\sqrt{\kappa_j} c_0(t)dt, ~~~ j=1,\ldots,n-1, 
\\
dB_{\rm out}(t) 
&=&
\sqrt{\gamma}c_0(t)+dB(t),  
\label{eq:H_independent_oscillator_H}
\end{eqnarray}
%\begin{eqnarray}
%L &=&\sqrt{\gamma }\,c_{0},  \label{eq:L_independent_oscillator_L} \\
%H &=&\omega _{0}c_{0}^{\ast }c_{0}+\sum_{j=1}^{n-1}\omega _{j}c_{j}^{\ast
%}c_{j}+\sum_{j=1}^{n-1}\sqrt{\kappa _{j}}\left( c_{0}^{\ast
%}c_{j}+c_{j}^{\ast }c_{0}\right) ,  \label{eq:H_independent_oscillator_H}
%\end{eqnarray}
where 
\begin{equation}
\gamma \equiv \sum_{j=1}^{n}\gamma _{j},~~\omega _{0}\equiv \frac{1}{\gamma }
\sum_{j,k=1}^{n}\sqrt{\gamma _{j}\gamma _{k}}\omega _{jk},
\label{eq:omega_0}
\end{equation}
and the other parameters $\omega_j, \kappa_j$ ($j=1,\ldots, n-1$) are given
in the proof.
\end{theorem}

{\bf Proof.}~ 
Let $R$ be a unitary matrix whose first row is $R_{1j}=\sqrt{\gamma
_{j}/\gamma }$, ($j=1,\ldots n$). Set $b_{j}^{\prime }\equiv
\sum_{k=1}^{n}R_{jk}a_{k}$, $j=1,\ldots n$. We have $[b_{j}^{\prime
},b_{k}^{\prime \ast }]=\delta _{jk}$. Clearly $L=C_-\mathbf{a}=\sqrt{\gamma 
}b_{1}^{\prime }$ and $[L, b_j^{\prime \ast}]=0$ for $j=2,\ldots n$. Let us
apply a further unitary transformation $V$ of the form $V=\left[ 
\begin{array}{cc}
1 & 0_{n-1}^{\top } \\ 
0_{n-1} & \tilde{V}
\end{array}
\right] $ with $0_{n-1}$ the column vector of length $n-1$ with all zero
entries and $\tilde{V}$ unitary in $\mathbb{C}^{(n-1)\times (n-1)}$ to be
specified later. We set 
\begin{equation*}
\mathbf{c} = \left[ 
\begin{array}{c}
c_{0} \\ 
c_{1} \\ 
\vdots \\ 
c_{n-1}
\end{array}
\right] \equiv V\,\mathbf{b}^{\prime }=VR\,\mathbf{a}.
\end{equation*}
We have $L=\sqrt{\gamma }c_{0}$. The Hamiltonian takes the form $H=\mathbf{c}^{\dag }VR\Omega R^{\dag }V^{\dag }\mathbf{c}=\mathbf{c}^{\dag }\Omega
^{\prime }\mathbf{c}$, where 
\begin{equation*}
\Omega ^{\prime }\equiv \left[ 
\begin{array}{cc}
1 & 0_{n-1}^{\top } \\ 
0_{n-1} & \tilde{V}
\end{array}
\right] R\Omega _{-}R^{\dag }\left[ 
\begin{array}{cc}
1 & 0_{n-1}^{\top } \\ 
0_{n-1} & \tilde{V}^{\dag }
\end{array}
\right] .
\end{equation*}
As the matrix $\tilde{V}$ is still arbitrary except being unitary, we may
choose it to diagonalize the lower right $(n-1)\times (n-1)$ block of $R\Omega R^{\dag }$, and with this choice we obtain $\Omega ^{\prime }$ of
the form 
\begin{equation}
\Omega _{\text{IO}}\equiv \left[ 
\begin{array}{cccc}
\omega _{0} & \varepsilon _{1}^{\ast } & \cdots & \varepsilon _{n-1}^{\ast }
\\ 
\varepsilon _{1} & \omega _{1} &  & 0 \\ 
\vdots &  & \ddots &  \\ 
\varepsilon _{n-1} & 0 &  & \omega _{n-1}
\end{array}
\right] .
\end{equation}
It can be readily verified that $\omega _{0}=\frac{1}{\gamma }\sum_{jk=1}^{n}
\sqrt{\gamma _{j}\gamma _{k}}\omega _{jk}$. Set $T = VR$ and the overall
unitary transform is thus $\mathbf{c} = T\, \mathbf{a}$. Finally we may
absorb the phases of the $\varepsilon _{k}$ into the modes, so without loss
of generality we may assume that they are real and non-negative, say $\varepsilon _{k}\equiv \sqrt{\kappa _{k}}$.

By Proposition \ref{prop:minimal_C_Omega}  and Theorem \ref{thm:DF}, we have the following corollary:

\begin{corollary}
For the realization (\ref{eq:c})-(\ref{eq:H_independent_oscillator_H}) constructed in Proposition \ref{prop:independent_oscillator}, if $\kappa_j=0$, ($j=1,\ldots, n-1$,) then the mode $
c_j$ is neither controllable nor observable.
\end{corollary}

Because the two realizations, $G\sim(I,C_-,\Omega_-)$ with $C_-,\Omega_-$ defined in (\ref{eq:independent_original_osillator_parameters}) and that in (\ref{eq:c})-(\ref{eq:H_independent_oscillator_H}), are unitarily equivalent, they have the same transfer function. In what follows we derive their transfer function. 

The following lemma turns out to be useful.

\begin{lemma}\label{lem:matrix_rep1}
 We have the algebraic identity that 
\begin{eqnarray*}
\left( \left[ 
\begin{array}{cccc}
a_{0} & b_{1} & \cdots & b_{n} \\ 
b_{1} & a_{1} & \ddots & 0 \\ 
\vdots &  & \ddots &  \\ 
b_{n} & 0 &  & a_{n}
\end{array}
\right]^{-1}\right) _{\rm row \; 1, column \; 1} =\dfrac{1}{a_{0}-\sum_{k=1}^{n}
\dfrac{b_{k}^{2}}{a_{k}}},
\end{eqnarray*}
where $(X)_{\rm row \; 1, column \; 1}$ means the entry on the intersection of
the first row and first column of a constant matrix $X$.
\end{lemma}

The proof is given in the Appendix.

We are now ready to present the transfer function.

\begin{corollary}\label{prop:bus}
The SISO quantum linear passive system $G\sim (I,C_{-},\Omega _{-})$ with $C_-, \Omega_-$ defined in (\ref{eq:independent_original_osillator_parameters}) has a
transfer function of the form 
\begin{equation}  \label{eq:bus}
G\left( s\right) =1-\frac{\gamma }{s+\frac{1}{2}\gamma +i\omega
_{0}+\sum_{k=1}^{n-1}\frac{\kappa _{k}}{s+i\omega _{k}}} .
\end{equation}
\end{corollary}

The proof follows Theorem \ref{prop:independent_oscillator} and Lemma \ref{lem:matrix_rep1}.

{\it Remark 6.}  Theorem \ref{prop:independent_oscillator} gives an independent-oscillator
realization of a quantum linear passive system, Fig.~\ref{fig:rep1}.
Unfortunately, because the unitary matrices $V$ and $R$ used in the proof of
Theorem \ref{prop:independent_oscillator} are by no means unique, it is
unclear whether this realization is unique or not, that is, whether the
parameters $\omega _{i}$ and $\kappa _{j}$ are uniquely determined by the system
parameters $\gamma _{i}$ and $\omega _{jk}$ in (\ref{eq:independent_original_osillator_parameters}) or not. In Theorem \ref{thm:independent_mode_unique} to be given in Subsection \ref{sec:equivalence}, we show that the independent-oscillator realization is unique under the assumption of minimal realization.

\subsubsection{Chain-mode realization}\label{subset:chain}

In the subsection we present the chain-mode realization of SISO quantum linear passive systems.

Let $G\sim (I,C_{\text{min}},\Omega _{\text{min}})$ be a Hurwitz stable  SISO quantum linear system  with $n_{\text{min}}$ the number of system oscillators. We assume that  $\Omega _{\text{min}}$ is diagonal and the entries of $C_{\text{min}}$ are non-negative;
specifically, 
\begin{equation} \label{eq:minimal}
\mathbf{\bar{a}}=\left[ 
\begin{array}{c}
\bar{a}_{1} \\ 
\vdots \\ 
\bar{a}_{n_{\text{min}}}
\end{array}
\right] ,\; \Omega _{\text{min}}=\mathrm{diag}\left( \bar{\omega}_{1},\cdots ,
\bar{\omega}_{n_{\text{min}}}\right) ,\;C_{\text{min}}= \left[\sqrt{\bar{\gamma}_{1}}, \cdots, \sqrt{\bar{\gamma}_{n_{\text{min}}}} \ \right]. 
%C_{\text{min}}=\left[ 
%\begin{array}{c}
%\sqrt{\bar{\gamma}_{1}} \\ 
%\vdots \\ 
%\sqrt{\bar{\gamma}_{n_{\text{min}}}}%
%\end{array}%
%\right] .  \label{eq:minimal}
\end{equation}

{\it Remark 7.} 
Because the matrix $\Omega_{\mathrm{min}} $ is Hermitian, it can always be diagonalized. Similarly by absorbing phases into system oscillators if necessary, the entries of the matrix $C_{\mathrm{min}}$ can be taken to be 
non-negative. Thus, given a Hurwitz stable quantum linear passive system,  one can always unitarily transform it to another one corresponding to  (\ref{eq:minimal}). Moreover, by
Proposition \ref{prop:siso_minimal}, minimality requires that $\bar{\omega}_{j}
\neq \bar{\omega}_{k}$ if $j\neq k$, and  $\bar{\gamma}_{j}\neq 0$, $j=1,\ldots, n_{\mathrm{min}}$.

\begin{figure}[htbp]
\centering
\includegraphics[width=0.300\textwidth]{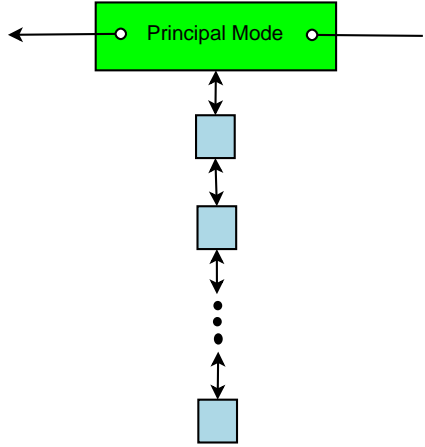}
\caption{The Chain-mode realization: the principal mode is coupled to a
non-damped mode which in turn is coupled to a finite chain of modes.}
\label{fig:repCM}
\end{figure}

In what follows we unitarily transform the system $G\sim (I,\Omega_{\rm min}, C_{\rm min})$ to a chain-mode realization of an assembly of interacting oscillators, Fig.~\ref{fig:repCM}.

\begin{theorem}\label{lem:continued_fraction}
 For the system $G\sim \left(I,C_{\rm
min},\Omega _{\rm min}\right) $ defined by  (\ref{eq:minimal}), there exists a
unitary transform $W$ such that the transformed modes 
\begin{equation} \label{eq:d}
\left[ 
\begin{array}{c}
\tilde{c}_{0} \\ 
\tilde{c}_{1} \\ 
\vdots \\ 
\tilde{c}_{n_{\mathrm{min}}-1}
\end{array}
\right] \equiv W\,\mathbf{\bar{a}}
\end{equation}
have the following realization:
\begin{eqnarray}
d\tilde{c}_0(t) 
&=&
-(\bar{\gamma}/2+i\tilde{\omega}_0) \tilde{c}_0(t)dt -i\sqrt{\tilde{\kappa}_1}\tilde{c}_1(t)dt -\sqrt{\bar{\gamma}}dB(t),
\label{eq:L_new}\\
d\tilde{c}_j(t)
&=&
-i\tilde{\omega}_j \tilde{c}_j(t)dt -i\sqrt{\tilde{\kappa}_j}\tilde{c}_{j-1}(t)dt-i\sqrt{\tilde{\kappa}_{j+1}}\tilde{c}_{j+1}(t)dt, ~~ j=1,\ldots, n_{\rm min}-2 ,
\\
d\tilde{c}_{n_{\rm min}-1}(t)
&=&
-i\tilde{\omega}_{n_{\rm min}-1} \tilde{c}_{n_{\rm min}-1}(t)dt -i\sqrt{\tilde{\kappa}_{n_{\rm min}-1}} \tilde{c}_{n_{\rm min}-2}(t)dt,
\\
dB_{\rm out}(t) 
&=&
\sqrt{\bar{\gamma}}\tilde{c}_0(t) dt + dB(t),
 \label{eq:H_new}
\end{eqnarray}
%
%\begin{eqnarray}
%L &=&\sqrt{\bar{\gamma}}\,d_{0},  \label{eq:L_new} \\
%H &=&\tilde{\omega}_{0}d_{0}^{\ast }d_{0}+\sum_{j=1}^{n_{\mathrm{min}}-1}
%\tilde{\omega}_{j}d_{j}^{\ast }d_{j}+\sqrt{\tilde{\kappa}_{0}}\left(
%d_{0}^{\ast }d_{1}+d_{0}d_{1}^{\ast }\right) +\sum_{j=1}^{n_{\mathrm{min}}-2}
%\sqrt{\tilde{\kappa}_{j}}\left( d_{j}^{\ast }d_{j+1}+d_{j}d_{j+1}^{\ast
%}\right) ,  \label{eq:H_new}
%\end{eqnarray}
where the parameters $\tilde{\omega}_{j}$ and $\tilde{\kappa}_{j}$ are given
respectively in (\ref{eq:tilde_omega_k}) and (\ref{eq:tilde_kappa_k}) in the
proof.
\end{theorem}

The proof is given in the Appendix.

{\it Remark 8.} 
In the literature of continued fraction, \cite{Wall}; \cite{WG}; \cite{HCB09}; \cite{WGC+12}, etc., the matrix 
\begin{equation*}
\mathcal{J} \equiv \left[ 
\begin{array}{cccccc}
\tilde{\omega} _{0} & \sqrt{\tilde{\kappa}_{1}} & 0 & \cdots & 0 & 0 \\ 
\sqrt{\tilde{\kappa}_{1}} & \tilde{\omega}_{1} & \sqrt{\tilde{\kappa}_{2}} & 
& 0 & 0 \\ 
0 & \sqrt{\tilde{\kappa}_{2}} & \tilde{\omega}_{2} & \ddots &  & \vdots \\ 
\vdots &  & \ddots & \ddots & \sqrt{\tilde{\kappa}_{n_{\mathrm{min}}-2}} & 0
\\ 
0 & 0 &  & \sqrt{\tilde{\kappa}_{n_{\mathrm{min}}-2}} & \tilde{\omega}_{n_{\mathrm{min}}-2} & \sqrt{\tilde{\kappa}_{n_{\mathrm{min}}-1}} \\ 
0 & 0 & \cdots & 0 & \sqrt{\tilde{\kappa}_{n_{\mathrm{min }}-1}} & \tilde{\omega}_{n_{\mathrm{min}}-1}
\end{array}
\right]
\end{equation*}
is often called a {\it Jacob matrix}. Clearly $\mathcal{J}$ is actually the Hamiltonian
matrix for the new system corresponding to the realization (\ref{eq:L_new})-(\ref{eq:H_new}).

Because the two realizations, $G\sim(I,C_{\rm min},\Omega_{\rm min})$ defined by (\ref{eq:minimal}) and that in (\ref{eq:L_new})-(\ref{eq:H_new}), are unitarily equivalent, they share the same transfer function. Next we study their transfer function.

We begin with the following lemma.

\begin{lemma}\label{lem:matrix_rep2}
 We have the algebraic identity that 
\begin{equation*}
\left[ 
\begin{array}{cccc}
a_{0} & b_{1} &  & 0 \\ 
b_{1} & a_{1} & \ddots &  \\ 
& \ddots & \ddots & b_{n} \\ 
0 &  & b_{n} & a_{n}
\end{array}
\right] _{\rm row \; 1, column \; 1}^{-1}=\dfrac{1}{a_{0}-\dfrac{b_{1}^{2}}{a_{1}-\dfrac{b_{2}^{2}}{
\begin{array}{ccc}
a_{2}- & \ddots &  \\ 
&  & -\dfrac{b_{n-1}^{2}}{a_{n-1}-\dfrac{b_{n}^{2}}{a_{n}}}
\end{array}
}}},
\end{equation*}
where $(X)_{\rm row \; 1, column \; 1}$ means the entry on the intersection of
the first row and first column of a constant matrix $X$.
\end{lemma}

The proof is given in the Appendix.

Based on Theorem \ref{lem:continued_fraction} and Lemma \ref{lem:matrix_rep2}, we may derive the transfer function.

\begin{corollary}\label{prop:chain_mode}
The SISO quantum linear
passive system $G\sim \left(I,C_{\rm min},\Omega _{\rm min}\right) $ has a transfer function in the
form of the continued fraction expansion 
\begin{eqnarray}
G\left( s\right)  
&=&I-\dfrac{\bar{\gamma}}{s+\dfrac{1}{2}\bar{\gamma}+i\omega _{0}+\dfrac{\tilde{\kappa}_{1}}{
\begin{array}{ccc}
s+i\tilde{\omega}_{1}+ & \ddots &  \\ 
&  & +\dfrac{\tilde{\kappa}_{n_{\mathrm{min}}-2}}{s+i\tilde{\omega}_{n_{\mathrm{min}}-2}+\dfrac{\tilde{\kappa}_{n_{\mathrm{min}}-1}}{s+i\tilde{\omega}_{n_{\mathrm{min}}-1}}}
\end{array}
}}. \label{eq:chain} 
\end{eqnarray}
%\begin{eqnarray}
%&&G\left( s\right)  \label{eq:chain} \\
%&=&S-\dfrac{\bar{\gamma}}{s+\dfrac{1}{2}\bar{\gamma}+i\omega _{0}+\dfrac{\tilde{\kappa}_{1}}{
%\begin{array}{ccc}
%s+i\tilde{\omega}_{1}+ & \ddots &  \\ 
%&  & +\dfrac{\tilde{\kappa}_{n_{\mathrm{min}}-2}}{s+i\tilde{\omega}_{n_{\mathrm{min}}-2}+\dfrac{\tilde{\kappa}_{n_{\mathrm{min}}-1}}{s+i\tilde{\omega}_{n_{\mathrm{min}}-1}}}
%\end{array}
%}}S.  \notag
%\end{eqnarray}
\end{corollary}

\subsubsection{Uniqueness of the independent-oscillator realization} \label{sec:equivalence}

 In Subsection \ref{sec:IO} an independent-oscillator
realization for SISO quantum linear passive systems is proposed. From the
construction it is unclear whether the parameters in this
independent-oscillator realization are unique,  Remark 6. In this
subsection we show that they are indeed unique if  minimality is assumed.

\begin{theorem}\label{thm:independent_mode_unique}
 Given a minimal quantum linear passive
system $G \sim (I,C_{\mathrm{min}},\Omega_{\mathrm{min}})$ in (\ref{eq:minimal}), its unitarily equivalent independent-oscillator realization is unique.
\end{theorem}

{\bf Proof.}~ 
Firstly, for the minimal
realization $G \sim (I,C_{\mathrm{min}},\Omega_{\mathrm{min}})$ in (\ref{eq:minimal}), by (\ref{eq:omega_0}) and (\ref{eq:tilde_omega_0}), $\omega_0
= \tilde{\omega}_0$. Secondly, by (\ref{eq:bus}) and (\ref{eq:chain}) we see
the transfer function takes the form 
\begin{equation}  \label{eq:G-Delta_a}
G(s) = 1-\frac{\gamma}{s+\frac{\gamma}{2}+i\omega_0 + \Delta(s)},
\end{equation}
where 
\begin{eqnarray}
\Delta (s) &\equiv&\sum_{k=1}^{n_{\mathrm{min}}-1}\dfrac{\kappa _{k}}{s
+i\omega_{k}}  \label{eq:Delta1_b} \\
&=&\dfrac{\tilde{\kappa}_{1}}{s +i\tilde{\omega}_{1}+\dfrac{\tilde{\kappa}_{2}}{
\begin{array}{ccc}
s+i\tilde{\omega}_{2}+ & \ddots &  \\ 
&  & +\dfrac{\tilde{\kappa}_{n_{\mathrm{min}}-2}}{s +i\tilde{\omega}_{n_{\mathrm{min}}-2}+\dfrac{\tilde{\kappa}_{n_{\mathrm{min}}-1}}{s +\tilde{\omega}_{n_{\mathrm{min}}-1}}}
\end{array}
}}  \label{eq:Delta2_b}
\end{eqnarray}
in the independent-oscillator and chain-mode realizations respectively.
Replacing $s$ with $i\omega$ in (\ref{eq:G-Delta_a}), (\ref{eq:Delta1_b})
and (\ref{eq:Delta2_b}) we have
\begin{equation}
G\left( i\omega \right) 
=
 1+\dfrac{i\gamma }{\omega +\omega _{0}-\dfrac{\gamma }{2}i - \hat{\Delta} \left( \omega \right) },  
 \label{eq:G-Delta}
\end{equation}
where 
\begin{eqnarray}
\hat{\Delta} \left( \omega \right) \equiv i\Delta(i\omega) &=&
\sum_{k=1}^{n_{\mathrm{min}}-1}\dfrac{\kappa _{k}}{\omega +\omega _{k}}
\label{eq:Delta1} \\
&=&\dfrac{\tilde{\kappa}_{1}}{\omega +\tilde{\omega}_{1}-\dfrac{\tilde{\kappa}_{2}}{
\begin{array}{ccc}
\omega+\tilde{\omega}_{2}- & \ddots &  \\ 
&  & -\dfrac{\tilde{\kappa}_{n_{\mathrm{min}}-2}}{\omega +\tilde{\omega}_{n_{\mathrm{min}}-2}-\dfrac{\tilde{\kappa}_{n_{\mathrm{min}}-1}}{\omega +\tilde{\omega}_{n_{\mathrm{min}}-1}}}
\end{array}
}}  \label{eq:Delta2}
\end{eqnarray}
in the independent-oscillator and chain-mode realizations respectively.
By Theorem \ref{lem:continued_fraction}, $\tilde{\omega}_j$ and $\tilde{\kappa}_j$ in (\ref{eq:Delta2}) are uniquely determined by  $C_{\mathrm{min}}$ and $\Omega_{\mathrm{min}}$,
that is, $\hat{\Delta} \left( \omega \right) $ is unique. On the other hand,
because $G=(I,C_{\mathrm{min}}, \Omega_{\mathrm{min}})$ is minimal, in (\ref{eq:Delta1}) $\omega_j \neq \omega_k$ if $j\neq k$, and $\kappa_i\neq 0$.
Clearly, for this single pole fraction form of $\hat{\Delta}
\left( \omega \right)$ in (\ref{eq:Delta1}), $\kappa_k$ and $\omega_k$ are unique. The proof is completed.

We notice that (\ref{eq:G-Delta_a}) implies that 
\begin{equation}  \label{eq:Sigma_Delta}
\Sigma \left( s\right) = \dfrac{1}{2}\dfrac{\gamma }{s+i\omega _{0}+\Delta
\left(s\right) }
\end{equation}
with $\Delta(s)$ given by (\ref{eq:Delta2_b}).

{\it Remark 9.} 
Given $\Delta(s)$ in (\ref{eq:Delta1_b}) and (\ref{eq:Delta2_b}), by (\ref{eq:Sigma_Delta}) an explicit form of $\Sigma(s)$ can be constructed, subsequently a quantum linear passive system $G(s) =
(I-\Sigma(s))((I+\Sigma(s)))^{-1}$ can be constructed. According to (\ref{eq:G-Delta_a}), $G(s)$ constructed in this way is always a genuine quantum system. In this sense, (\ref{eq:Sigma_Delta}) indicates what
type of lossless positive real functions can generate a quantum
linear passive system (which is lossless bounded real).

%%%%%%%%%%%%%%%%%%%%%%%%%%%%%%%%
%%%%%%%%%%%%%%%%%%%%%%%%%%%%%%%%
%%%%%%%%%%%%%%%%%%%%%%%%%%%%%%%%
\section{Conclusion}\label{sec:conclusion}
In this paper we have studied the realization theory of quantum linear systems. We have shown the equivalence between controllability and observability of general quantum linear systems, and in particular in the passive case they are equivalent to Hurwitz stability. Based on controllability and observability, a special form of realization has been proposed for general quantum linear systems which can be regarded as the complex-domain counterpart of the so-called decoherence-free subspace decomposition studied in \cite{NY13}. Specific to quantum linear passive systems,  formulas for calculating the cardinality of minimal realizaitons are proposed.  A specific realization is proposed for the multi-input-multi-output case which is closely related to controllability and observability decomposition. Finally, two realizations,  the independent-oscillator realization and the chain-mode realization, have been derived for the single-input-single-output case. It is expected that these results will find applications in quantum systems design.

%%%%%%%%%%%%%%%%%%%%%%%%%%%%%%%%
%%%%%%%%%%%%%%%%%%%%%%%%%%%%%%%%
%%%%%%%%%%%%%%%%%%%%%%%%%%%%%%%%

\section*{Acknowledgment}

The authors wish to thank Daniel Burgarth for pointing out Reference 
\cite{WGC+12}. The second author would like to thank Runze Cai, Lei Cui and Zhiyang Dong for helpful discussions. 

%%%%%%%%%%%%%%%%%%%%%%%%%%%%%%%%
%%%%%%%%%%%%%%%%%%%%%%%%%%%%%%%%
%%%%%%%%%%%%%%%%%%%%%%%%%%%%%%%%
%\appendix
%
%\section*{Proofs.}

\noindent {\bf Appendix.}

{\it Proof of Lemma \ref{lem:ctrb_obsv}.} We first show that the dimension of the space $\mathrm{Ker}\left( \mathbf{O}_{s}\right) \mathbf{\cap }\mathrm{Ker}\left( \mathbf{O}_{s}J_{n}\right)$ is even. If a nonzero vector
\begin{equation}
v=\left[ 
\begin{array}{c}
v_{1} \\ 
v_{2}
\end{array}
\right] \in \mathrm{Ker}\left( \mathbf{O}_{s}\right) \mathbf{\cap }\mathrm{Ker}\left( \mathbf{O}_{s}J_{n}\right)  \label{eq:feb28_4}
\end{equation}
with $v_1,v_2\in \mathbb{C}^n$, then 
\begin{equation}
\mathcal{C}\left[ 
\begin{array}{c}
v_{1} \\ 
v_{2}
\end{array}
\right] =\left[ 
\begin{array}{c}
C_{-}v_{1}+C_{+}v_{2} \\ 
C_{+}^{\#}v_{1}+C_{-}^{\#}v_{2}
\end{array}
\right] =0, ~~ \mathcal{C}J_{n}\left[ 
\begin{array}{c}
v_{1} \\ 
v_{2}
\end{array}
\right] =\left[ 
\begin{array}{c}
C_{-}v_{1}-C_{+}v_{2} \\ 
C_{+}^{\#}v_{1}-C_{-}^{\#}v_{2}
\end{array}
\right] =0,  \label{eq:feb28_temp3}
\end{equation}
which are equivalent to
\begin{equation*}
\mathcal{C}\left[ 
\begin{array}{cc}
v_{1} & 0 \\ 
0 & v_{2}
\end{array}
\right] =0.
\end{equation*}
That is,
\begin{equation}
\mathcal{C}\left[ 
\begin{array}{c}
v_{1} \\ 
v_{2}
\end{array}
\right] =0, ~~  \mathcal{C}J_{n}\left[ 
\begin{array}{c}
v_{1} \\ 
v_{2}
\end{array}
\right] =0  \Longleftrightarrow \mathcal{C}\left[ 
\begin{array}{c}
v_{1} \\ 
0
\end{array}
\right] =0, ~~ \mathcal{C}\left[ 
\begin{array}{c}
0 \\ 
v_{2}
\end{array}
\right] =0.  \label{eq:feb28_3}
\end{equation}
On the other other hand, by (\ref{eq:feb28_4}) we also have  
\begin{equation}
\mathcal{C}J_{n}\Omega \left[ 
\begin{array}{c}
v_{1} \\ 
v_{2}
\end{array}
\right] =0, ~~~ \mathcal{C}J_{n}\Omega J_n\left[ 
\begin{array}{c}
v_{1} \\ 
v_{2}
\end{array}
\right] =0,  \label{eq:feb28_temp8}
\end{equation}
which are equivalent to
\begin{equation*}
\mathcal{C}J_{n}\Omega \left[ 
\begin{array}{cc}
v_{1} & 0 \\ 
0 & v_{2}
\end{array}
\right] =0.
\end{equation*}
Therefore we have
\begin{equation}
\mathcal{C}J_{n}\Omega \left[ 
\begin{array}{c}
v_{1} \\ 
v_{2}
\end{array}
\right] =0, ~~ \mathcal{C}J_{n}\Omega J_n\left[ 
\begin{array}{c}
v_{1} \\ 
v_{2}
\end{array}
\right] =0\Longleftrightarrow \mathcal{C}J_{n}\Omega \left[ 
\begin{array}{c}
v_{1} \\ 
0
\end{array}
\right] =0, ~~ \mathcal{C}J_{n}\Omega \left[ 
\begin{array}{c}
0 \\ 
v_{2}
\end{array}
\right] =0.  \label{eq:feb28_5}
\end{equation}
Analogously it can be shown that
\begin{equation}
\mathcal{C}(J_{n}\Omega )^{k}\left[ 
\begin{array}{c}
v_{1} \\ 
v_{2}
\end{array}
\right]=0, ~~ \mathcal{C}J_n(J_{n}\Omega )^{k}\left[ 
\begin{array}{c}
v_{1} \\ 
v_{2}
\end{array}
\right]=0\Longleftrightarrow \mathcal{C}(J_{n}\Omega
)^{k}\left[ 
\begin{array}{c}
v_{1} \\ 
0
\end{array}
\right] =0, ~~ \mathcal{C}(J_{n}\Omega )^{k}\left[ 
\begin{array}{c}
0 \\ 
v_{2}
\end{array}
\right] =0, ~~ k\geq 1  \label{eq:feb28_6}
\end{equation}
(\ref{eq:feb28_3}), (\ref{eq:feb28_5}) and (\ref{eq:feb28_6}) indicate that 
\begin{equation}
v=\left[ 
\begin{array}{c}
v_{1} \\ 
v_{2}
\end{array}
\right] \in \mathrm{Ker}\left( \mathbf{O}_{s}\right) \mathbf{\cap }\mathrm{Ker}\left( \mathbf{O}_{s}\mathbf{J}_{n}\right) \Longleftrightarrow \left[ 
\begin{array}{c}
v_{1} \\ 
0
\end{array}
\right] , ~ \left[ 
\begin{array}{c}
0 \\ 
v_{2}
\end{array}
\right] \in \mathrm{Ker}\left( \mathbf{O}_s\right) \mathbf{\cap }\mathrm{Ker}\left( \mathbf{O}_s J_{n}\right) \mathbf{.}  \label{eq:feb28_7}
\end{equation}
Moreover, it can be readily shown that
\begin{eqnarray}
\left[ 
\begin{array}{c}
v_{1} \\ 
0
\end{array}
\right]  \in \mathrm{Ker}\left( \mathbf{O}_s\right) \mathbf{\cap }\mathrm{Ker}\left( \mathbf{O}_s J_{n}\right) &\Longleftrightarrow&  \left[ 
\begin{array}{c}
0 \\ 
v_{1}^{\#}
\end{array}
\right] \in \mathrm{Ker}\left( \mathbf{O}_{s}\right) \mathbf{\cap }\mathrm{Ker}\left( \mathbf{O}_{s}J_{n}\right) \mathbf{,}  \label{eq:feb28_8} \\
\left[ 
\begin{array}{c}
0 \\ 
v_{2}
\end{array}
\right]  \in \mathrm{Ker}\left( \mathbf{O}_{s}\right) \mathbf{\cap }
\mathrm{Ker}\left( \mathbf{O}_{s}J_{n}\right)  &\Longleftrightarrow& \left[ 
\begin{array}{c}
v_{2}^{\#} \\ 
0
\end{array}
\right] \in \mathrm{Ker}\left( \mathbf{O}_s\right) \mathbf{\cap }\mathrm{Ker}\left( \mathbf{O}_s J_{n}\right) \mathbf{,}  \label{eq:feb28_9}
\end{eqnarray}
As a result, one can choose an orthonormal basis of $\mathrm{Ker}\left( \mathbf{O}_{s}\right) \mathbf{\cap }\mathrm{Ker}\left( \mathbf{O}_{s}J_{n}\right) $ to be one of the form
\begin{equation*}
\left[ 
\begin{array}{c}
v_{1} \\ 
0
\end{array}
\right] , ~ \left[ 
\begin{array}{c}
0 \\ 
v_{1}^{\#}
\end{array}
\right] ,\cdots ,\left[ 
\begin{array}{c}
v_{l} \\ 
0
\end{array}
\right] , ~ \left[ 
\begin{array}{c}
0 \\ 
v_{l}^{\#}
\end{array}
\right] .
\end{equation*}
Therefore, the dimension of the space $\mathrm{Ker}\left( \mathbf{O}_{s}\right) \mathbf{\cap }\mathrm{Ker}\left( \mathbf{O}_{s}J_{n}\right)$ is even. Here we take it to be $2l$. 

Secondly, we construct $V_1 \in \mathbb{C}^{2n\times 2l}$. 
Noticing $\mathrm{Ker}\left( \mathbf{O}_{s}J_{n}[
\begin{array}{cc}
I_{n} & 0_{n}
\end{array}
]^{T}\right) =\mathrm{Ker}\left( \mathbf{O}_{s}[
\begin{array}{cc}
I_{n} & 0_{n}
\end{array}
]^{T}\right) $, we have
\[
\left[
\begin{array}{c}
v_i \\
0
\end{array}
\right]
 \in  \mathrm{Ker}\left( \mathbf{O}_{s}\right) \mathbf{\cap }\mathrm{Ker}\left( \mathbf{O}_{s}J_{n}\right)
\Longleftrightarrow
v_i \in \mathrm{Ker}\left( \mathbf{O}_{s}\left[
\begin{array}{c}
I_{n} \\
 0_{n}
\end{array}
\right]\right) .
\]
Thus it is sufficient to construct the orthonormal basis vectors $v_1, \ldots, v_l$ for
the space $\mathrm{Ker}\left( \mathbf{O}_{s}[
\begin{array}{cc}
I_{n} & 0_{n}
\end{array}
]^{T}\right) $.
This can be done by the Gram-Schmidt orthogonalisation procedure.  Define
\begin{equation}
V_{1}\equiv \left[ 
\begin{array}{cccccc}
v_{1} & \cdots  & v_{l} & 0 & \cdots  & 0 \\ 
0 & \cdots  & 0 & v_{1}^{\#} & \cdots  & v_{l}^{\#}
\end{array}
\right] \in \mathbb{C}^{2n\times 2l}.  \label{eq:V1}
\end{equation}
For the above construction, $\mathrm{Range}(V_{1})=\mathrm{Ker}\left( \mathbf{O}_{s}\right) \cap 
\mathrm{Ker}\left( \mathbf{O}_{s}J_{n}\right) $. (\ref{eq:W1_range}) is
established.

Thirdly, we construct the matrix $V_{2}$. If a normalized vector $v_{l+1}\in 
\mathbb{C}^{n}$ such that for all $k=1,\ldots ,l$, $v_{l+1}^{\dagger }v_{k}=0
$, then $(v_{l+1}^{\#})^{\dagger }v_{k}^{\#}=0$. That\ is, the normalized
vectors $\left[ 
\begin{array}{c}
v_{l+1} \\ 
0
\end{array}
\right] $ and $\left[ 
\begin{array}{c}
0 \\ 
v_{l+1}^{\#}
\end{array}
\right] $ are orthogonal to the space $\mathrm{Range}(V_{1})$. Of course 
$\left[ 
\begin{array}{c}
v_{l+1} \\ 
0
\end{array}
\right] $ and $\left[ 
\begin{array}{c}
0 \\ 
v_{l+1}^{\#}
\end{array}
\right] $ are orthogonal to each other too. By the Gram-Schmidt orthogonalisation
procedure an orthonormal basis $\left\{ v_{l+1},\ldots ,v_{n}\right\} $ can
be found for the orthogonal space of the space spanned by the vectors $
\left\{ v_{1},\ldots ,v_{l}\right\} .$ The an orthonormal matrix $V_{2}$ can
be constructed to be 
\begin{equation*}
V_{2}\equiv \left[ 
\begin{array}{cccccc}
v_{l+1} & \cdots  & v_{n} & 0 & \cdots  & 0 \\ 
0 & \cdots  & 0 & v_{l+1}^{\#} & \cdots  & v_{n}^{\#}
\end{array}
\right] \in \mathbb{C}^{2n\times 2(n-l)}.
\end{equation*}

Fourthly, define $V\equiv [ 
\begin{array}{cc}
V_{1} & V_{2}
\end{array}
] $. Clearly, $V^{\dagger }V=I_{2n}$ which establishes (\ref{eq:W_unitary}). 

Finally, because $V_{1}^{\dagger }J_{n}=J_{l}V_{1}^{\dagger }$,
we have 
\begin{equation*}
V^{\dagger }J_{n}V=\left[ 
\begin{array}{cc}
V_{1}^{\dagger }J_{n}V_{1} & V_{1}^{\dagger }J_{n}V_{2} \\ 
V_{2}^{\dagger }J_{n}V_{1} & V_{2}^{\dagger }J_{n}V_{2}
\end{array}
\right] =\left[ 
\begin{array}{cc}
J_{l} & 0 \\ 
0 & J_{n-l}
\end{array}
\right] ,
\end{equation*}
which is (\ref{eq:W_symplectic}). The proof is completed.

%%%%%%%%%%%%%%%%%%%%%%%%%%%%%%%
\textit{Proof of Proposition \ref{prop:siso_minimal}. } Without loss of
generality, assume that $\Omega _{-}$ is diagonal. (Otherwise, there exists
a unitary matrix $T$ such that $\bar{\Omega}=T\Omega _{-}T^{\dag }$ is
diagonal. Correspondingly, denote $\bar{P}_{\omega }=TP_{\omega }T^{\dag }$
and $\bar{C}=C_{-}T^{\dag }$. Then $\bar{C}\bar{P}_{\omega }\bar{C}^{\dag
}=C_{-}P_{\omega }C^{\dag }$.) Let there be $r$ non-zero entries in the row
vector $C_{-}$. Because $\Omega _{-}$ is diagonal, if the $i$th element of $C_{-}$ is zero, then the $i$th column of the matrix in (\ref{eq:Os}) is a
zero column. As a result, for minimality we need only consider non-zero
elements of $C_{-}$. Without loss of generality, assume $C_{-}=[C_{1}~~0]$,
where $C_{1}=[c_{1}~c_{2}~\cdots ~c_{r}]$ with $c_{i}\neq 0$, $(i=1,\ldots
,r)$. Correspondingly, partition $\Omega_- $ as 
\begin{equation*}
\Omega _{-}=\left[ 
\begin{array}{cc}
\Omega _{1} & 0 \\ 
0 & \Omega _{2}
\end{array}
\right] ,
\end{equation*}
where $\Omega _{1}$ is a $r\times r$ square diagonal matrix with $\omega
_{1},\ldots ,\omega _{r}$ being diagonal entries. Clearly, 
\begin{equation}
\mathrm{rank}\left[ 
\begin{array}{c}
C_{-} \\ 
C_{-}\Omega _{-} \\ 
\ \vdots \\ 
C_{-}\Omega _{-}^{n-1}
\end{array}
\right] =\mathrm{rank}\left[ 
\begin{array}{c}
C_{1} \\ 
C_{1}\Omega _{1} \\ 
\ \vdots \\ 
C_{1}\Omega _{1}^{r-1}
\end{array}
\right] .  \label{eq:rank_siso_1}
\end{equation}
Notice that 
\begin{equation}
\left[ 
\begin{array}{c}
C_{1} \\ 
C_{1}\Omega _{1} \\ 
\vdots \\ 
C_{1}\Omega _{1}^{r-1}
\end{array}
\right] =\left[ 
\begin{array}{cccc}
1 & 1 & \cdots & 1 \\ 
\omega _{1} & \omega _{2} & \cdots & \omega _{r} \\ 
\vdots & \vdots & \vdots & \vdots \\ 
\omega _{1}^{r-1} & \omega _{2}^{r-1} & \cdots & \omega _{r}^{r-1}
\end{array}
\right] \left[ 
\begin{array}{cccc}
c_{1} &  &  &  \\ 
& c_{2} &  &  \\ 
&  & \ddots &  \\ 
&  &  & c_{r}
\end{array}
\right] .
\end{equation}
According to Lemma \ref{thm:stab_obsr_cont} and noticing $c_{i}\neq 0$ for $i=1,\ldots ,r$, 
\begin{equation}
n_{\mathrm{min}}=\mathrm{rank}\left[ 
\begin{array}{c}
C_{1} \\ 
C_{1}\Omega _{1} \\ 
\vdots \\ 
C_{1}\Omega _{1}^{r-1}
\end{array}
\right] =\mathrm{rank}\left[ 
\begin{array}{cccc}
1 & 1 & \cdots & 1 \\ 
\omega _{1} & \omega _{2} & \cdots & \omega _{r} \\ 
\vdots & \vdots & \vdots & \vdots \\ 
\omega _{1}^{r-1} & \omega _{2}^{r-1} & \cdots & \omega _{r}^{r-1}
\end{array}
\right] .
\end{equation}
Let $\ell $ be the total number of distinct diagonal entries of the matrix $\Omega _{1}$. By a property of the Vandermonde matrices, $\ell =n_{\mathrm{min}}$. Finally, denote the distinct eigenvalues of $\Omega _{1}$ by $\hat{\omega}_{1},\ldots ,\hat{\omega}_{\ell }$. For each $i=1,\ldots ,\ell $,
because $c_{i}\neq 0$, $C_{-}P_{\hat{\omega}_{i}}C_{-}^{\dag }\neq 0$. So we
have shown that the number $n_{\text{min}}$ of system oscillators of a minimal
realization $\left( S,C_{\text{min}},\Omega _{\text{min}}\right) $ equals the
total number of elements of the set $\sigma (\Omega _{-},C_{-})$ defined in (\ref{eq:spectral}).

%%%%%%%%%%%%%%%%%%%%%%%%%%%%%%%
\textit{Proof of Theorem \ref{thm:MIMO}.} The proof can be done by
construction. Let $\mathrm{rank}(C_{-})=r>0$. Firstly, according to \cite[Theorem 5.6.4]{Bernstein09} there exist unitary matrices $R_{1}\in \mathbb{C}^{m\times m}$ and $R_{2}\in \mathbb{C}^{n\times n}$ such that 
\begin{equation}
R_{1}C_{-}R_{2}=\left[ 
\begin{array}{cc}
\sigma (C_{-})_{r\times r} & 0 \\ 
0 & 0
\end{array}
\right]
\end{equation}
where $\sigma (C_{-})$ is a diagonal matrix with diagonal entries being
singular values of the matrix $C_-$. Partition the matrix $R_{2}^{\dag
}\Omega _{-}R_{2}$ accordingly, and denote 
\begin{equation}
\bar{\Omega} = \left[ 
\begin{array}{cc}
\tilde{\Omega}_{1} & \tilde{\Omega}_{2} \\ 
\tilde{\Omega}_{2}^{\dag } & \tilde{\Omega}_{3}
\end{array}
\right] \equiv R_{2}^{\dag }\Omega _{-}R_{2}.
\end{equation}
Define the unitary transformations
\begin{equation}
\left[ 
\begin{array}{c}
\tilde{b}_{\mathrm{in,pr}}(t) \\ 
\tilde{b}_{\mathrm{in,aux}}(t)
\end{array}
\right] \equiv R_{1}b(t),~~\left[ 
\begin{array}{c}
\tilde{b}_{\mathrm{out,pr}}(t) \\ 
\tilde{b}_{\mathrm{out,aux}}(t)
\end{array}
\right] \equiv R_{1}b_{\mathrm{out}}(t),~~\left[ 
\begin{array}{c}
\tilde{a}_{\mathrm{pr}}(t) \\ 
a_{aux}(t)
\end{array}
\right] \equiv R_{2}^{\dag }a(t),
\end{equation}
where all the first blocks on the left-hand side are a row vector of
dimension $r$. Then $G$ is unitarily equivalent to the following system 
\begin{eqnarray}
\dot{\tilde{a}}_{\mathrm{pr}} 
&=&
-(\sigma (C_{-})^{2}/2+i\tilde{\Omega}_{1})\tilde{a}_{\mathrm{pr}}-i\tilde{\Omega}_{2}a_{\mathrm{aux}}-\sigma (C_{-})\tilde{b}_{\mathrm{in,pr}}(t),  
\label{eq:system2_a} \\
\dot{a}_{\mathrm{aux}} 
&=&
-i\tilde{\Omega}_{2}^{\dag }\tilde{a}_{\mathrm{pr}}-i\tilde{\Omega}_{3}a_{\mathrm{aux}},  
\label{eq:system2_b} \\
\tilde{b}_{\mathrm{out,pr}} 
&=&
\sigma (C_{-})\tilde{a}_{\mathrm{pr}}+\tilde{b}_{\mathrm{in,pr}}(t), 
 \label{eq:system2_c} \\
\tilde{b}_{\mathrm{out,aux}} 
&=&
\tilde{b}_{\mathrm{in,aux}}(t).
\label{eq:system2_d}
\end{eqnarray}
By Schur decomposition there exists a unitary matrix $T\in \mathbb{C}^{(n-r)\times (n-r)}$ such that 
\begin{equation}
\tilde{\Omega}_{3}=T\left[ 
\begin{array}{cc}
\sigma (\tilde{\Omega}_{3}) & 0 \\ 
0 & 0
\end{array}
\right] T^{\dag }.
\end{equation}
Accordingly, denote $[\tilde{\Omega}_{21}~~\tilde{\Omega}_{22}]\equiv \tilde{\Omega}_{2}T^{\dag }.$ As a result, applying the unitary transformation 
\begin{equation}
\left[ 
\begin{array}{c}
\tilde{a}_{\mathrm{pr}} \\ 
\tilde{a}_{\mathrm{aux},1} \\ 
\tilde{a}_{\mathrm{aux},2}
\end{array}
\right] \equiv \left[ 
\begin{array}{cc}
I_{r\times r} & 0 \\ 
0 & T
\end{array}
\right] \left[ 
\begin{array}{c}
\tilde{a}_{\mathrm{pr}} \\ 
a_{\mathrm{aux}}
\end{array}
\right]
\end{equation}
to (\ref{eq:system2_a})-(\ref{eq:system2_b}) yields the final realization (\ref{eq:system3_a})-(\ref{eq:system3_e}). Clearly the realization (\ref{eq:system3_a})-(\ref{eq:system3_e}) corresponds to a quantum linear passive system whose the parameters are given in (\ref{eq:C_Omega}).
%Finally, note that if $\tilde{\Omega}_{21}=0$, then $\tilde{a}_{\mathrm{aux},1}$ is an uncontrollable mode; likewise, if $\tilde{\Omega}_{22}=0$, then $\tilde{a}_{\mathrm{aux},2}$ is an uncontrollable mode. However. if the
%system is Hurwitz stable, by Lemma \ref{thm:stab_obsr_cont}, it is both
%controllable and observable. As a result, if the system is Hurwitz stable, $\tilde{\Omega}_{21}\neq 0$, and $\tilde{\Omega}_{22}\neq 0$.

%%%%%%%%%%%%%%%%%%%%%%%%%%%%%%%
\textit{Proof of Lemma \ref{lem:matrix_rep1}.} We show this by induction. It
is clear true for $n=1$, so we the assume it is true for a given $n$ and
establish for $n+1$. Let us write $E_{11}\left( M\right) $ for the first
entry (row 1, column 1) of a matrix $M$. Let us consider a sequence 
\begin{equation*}
M_{n}=\left[ 
\begin{array}{cccc}
a_{0} & b_{1} & \cdots & b_{n} \\ 
b_{1} & a_{1} & \ddots & 0 \\ 
\vdots &  & \ddots &  \\ 
b_{n} & 0 &  & a_{n}
\end{array}
\right]
\end{equation*}
of matrices, then 
\begin{equation*}
M_{n+1}\equiv \left[ 
\begin{array}{cc}
M_{n} & b_{n+1}e_{n} \\ 
b_{n+1}e_{n}^{\top } & a_{n+1}
\end{array}
\right] ,\text{ where }e_{n}=\left[ 
\begin{array}{c}
1 \\ 
0 \\ 
\vdots \\ 
0
\end{array}
\right] \in \mathbb{C}^{n+1}.
\end{equation*}
We recall the Schur-Feshbach inversion formula for a matrix in block form 
\begin{equation} \label{eq:mar18_appendix_1}
\left[ 
\begin{array}{cc}
A_{11} & A_{12} \\ 
A_{21} & A_{22}
\end{array}
\right] ^{-1}=\left[ 
\begin{array}{cc}
Y^{-1} & -Y^{-1}A_{12}A_{22}^{-1} \\ 
-A_{22}^{-1}A_{21}Y^{-1} & 
A_{22}^{-1}+A_{22}^{-1}A_{21}Y^{-1}A_{12}A_{22}^{-2}
\end{array}
\right]
\end{equation}
where $Y=A_{11}-A_{21}A_{22}^{-1}A_{21}$. From the Schur-Feshbach formula we
deduce that 
\begin{equation*}
E_{11}\left( M_{n+1}^{-1}\right) =E_{11}\left( (M_{n}-\dfrac{b_{n+1}^{2}}{%
a_{n+1}}e_{n}e_{n}^{\top })^{-1}\right) .
\end{equation*}
However, the matrix $M_{n}-(b_{n+1}^{2}/a_{n+1})e_{n}e_{n}^{\top }$ is
identical to $M_{n}$ except that we replace the first row first column entry 
$a_{0}$ with $a_{0}-(b_{n+1}^{2}/a_{n+1})$, and by assumption we should then
have 
\begin{equation*}
E_{11}\left( (M_{n}-\dfrac{b_{n+1}^{2}}{a_{n+1}}e_{n}e_{n}^{\top
})^{-1}\right) =\dfrac{1}{\left( a_{0}-\dfrac{b_{n+1}^{2}}{a_{n+1}}\right)
-\sum_{k=1}^{n}\dfrac{b_{k}^{2}}{a_{k}}}.
\end{equation*}
This establishes the formula for $n+1$, and so the formula is true by
induction. 

%%%%%%%%%%%%%%%%%%%%%%%%%%%%%%%
\textit{Proof of Proposition \ref{lem:continued_fraction}.}  The spectral distribution $\Phi $ associated with a
SISO system $G\sim \left( S,C_{-},\Omega _{-}\right) $ is defined through
the Stieltjes' integral, i.e., 
\begin{equation*}
\int_{-\infty }^{\infty }e^{it\omega }d\Phi \left( \omega \right) =\frac{1}{C_{-}C_{-}^{\dag }}C_{-}e^{it\Omega _{-}}C_{-}^{\dag },
\end{equation*}
where the normalization coefficient $C_{-}C_{-}^{\dag }>0$. In
particular, in terms of the specific minimal realization  $G\sim (S,C_{\mathrm{min}},\Omega _{\mathrm{min}}) $ given in (\ref{eq:minimal}), we have
\begin{equation}
d\Phi \left( \omega \right) =\sum_{j=1}^{n_{\text{min}}}\frac{\bar{\gamma}_{j}}{\bar{\gamma}}\delta \left( \omega -\bar{\omega}_{j}\right) \,d\omega
\equiv \bar{\mu}(\omega )d\omega ,  \label{eq:spectral_distribution_minimal}
\end{equation}
where 
\begin{equation}
\bar{\gamma}\equiv \sum_{j=1}^{n_{\mathrm{min}}}\bar{\gamma}_{j}.
\label{eq:bar_gamma}
\end{equation}
That is, the cardinality of the support of $d\Phi $ is exactly the number of oscillators $n_{\rm min}$ in the minimal
realization of $G\sim (S,C_{\mathrm{min}},\Omega _{\mathrm{min}}) $.
The spectral distribution defined in (\ref{eq:spectral_distribution_minimal})  has only finitely many point
supports. We define an inner product for polynomials in the field of
real numbers in terms of this discrete spectral distribution. More
specifically, given two real polynomials $P(\omega )$ and $Q(\omega )$,
define their inner product with respect to $\bar{\mu}$ to be 
\begin{equation}
\langle P,Q\rangle _{\bar{\mu}}\equiv \int_{-\infty }^{\infty }P(\omega
)Q(\omega )\bar{\mu}(\omega )d\omega =\sum_{j=1}^{n_{\mathrm{min}}}\frac{\bar{\gamma}_{j}}{\bar{\gamma}}P(\bar{\omega}_{j})Q(\bar{\omega}_{j}).
\end{equation}
The norm of a polynomial $P(\omega )$ is of course $\Vert P\Vert \equiv 
\sqrt{\langle P,P\rangle _{\bar{\mu}}}$. Next we introduce a sequence of $n_{\mathrm{min}}$ orthogonal polynomials $\{P_{i}\}$, which are defined via the
Gram-Schmidt orthogonalization: 
\begin{equation*}
P_{0}(\omega )\equiv 1,~
P_{j}(\omega )=\omega ^{j}-\sum_{k=0}^{j-1}\frac{\langle \omega ^{j},P_{k}\rangle _{\bar{\mu}}}{\langle P_{k},P_{k}\rangle_{\bar{\mu}}}P_{k}(\omega),~~
j=1,\ldots ,n_{\mathrm{min}}-1,
\end{equation*}
where $\langle \omega ^{j},P_{k}\rangle _{\bar{\mu}}$ is to be understood as 
$\langle \omega ^{j},P_{k}\rangle _{\bar{\mu}}=\int_{-\infty }^{\infty
}\omega ^{j}P_{k}(\omega )\bar{\mu}(\omega )d\omega .$ It is easy to verify
that the above orthogonal polynomial sequence $\{P_{j}\}_{j=0}^{n_{\mathrm{min}}}$ satisfies the following three-term recurrence relation, \cite[Theorem 1.27]{WG} 
\begin{equation}
P_{k+1}(\omega )=(\omega -\tilde{\omega}_{k})P_{k}(\omega )-\sqrt{\tilde{\kappa}}_{k}P_{k-1}(\omega ),~~
k=0,\ldots ,n_{\mathrm{min}}-1,
\end{equation}
where $\tilde{\kappa}_{0}\equiv \Vert P_{0}\Vert $ and the convention $P_{-1}\equiv 0$ is assumed. Clearly, 
\begin{equation}
\tilde{\omega}_{k}=\frac{\langle \omega P_{k},P_{k}\rangle _{\bar{\mu}}}{
\langle P_{k},P_{k}\rangle _{\bar{\mu}}},~~k=0,\ldots ,n_{\mathrm{min}}-1,
\label{eq:tilde_omega_k}
\end{equation}
and 
\begin{equation}
\tilde{\kappa}_{k}=\sqrt{\frac{\langle P_{k},P_{k}\rangle _{\bar{\mu}}}{\langle P_{k-1},P_{k-1}\rangle _{\bar{\mu}}}},~~
k=1,\ldots ,n_{\mathrm{min}}-1.  \label{eq:tilde_kappa_k}
\end{equation}
(Note that $\tilde{\kappa}_{k}\neq 0$, $k=0,\ldots ,n_{\mathrm{min}}-1$.)
According to (\ref{eq:tilde_omega_k}), we have 
\begin{equation}
\tilde{\omega}_{0}=\frac{1}{\bar{\gamma}}\sum_{j=1}^{n_{\mathrm{min}}}\bar{\gamma}_{j}\bar{\omega}_{j}.  \label{eq:tilde_omega_0}
\end{equation}
By normalizing $\{P_{j}\}_{j=0}^{n_{\mathrm{min}}}$, that is define $\tilde{P}_{j}\equiv \frac{1}{\Vert P_{j}\Vert }P_{j}$, we can get a set of
orthonormal polynomial sequence $\{\tilde{P}_{j}\}_{j=0}^{n_{\mathrm{min}}}$. We define a new set of oscillators to be
\begin{eqnarray}
\tilde{c}_{0} &\equiv &\sum_{j=1}^{n_{\mathrm{min}}}\sqrt{\frac{\bar{\gamma}_{j}}{\bar{\gamma}}}\tilde{P}_{0}(\bar{\omega}_{j})\bar{a}_{j},  \label{eq:b_chain}
\\
\tilde{c}_{k} &\equiv &\sum_{j=1}^{n_{\mathrm{min}}}\sqrt{\frac{\bar{\gamma}_{j}}{\bar{\gamma}}}\tilde{P}_{k}(\bar{\omega}_{j})\bar{a}_{j},~~~k=1,\ldots ,n_{\mathrm{min}}-1.  \label{eq:d_k_chain}
\end{eqnarray}
It can be verified that the transformation (\ref{eq:b_chain})-(\ref{eq:d_k_chain}) is unitary. Moreover, 
\begin{equation}
\tilde{c}_{0}=\frac{1}{\sqrt{\bar{\gamma}}}\sum_{j=1}^{n_{\mathrm{min}}}\sqrt{\bar{\gamma}_{j}}\bar{a}_{j},  \label{eq:b_2}
\end{equation}
and the canonical commutation relations $[\tilde{c}_{0},\tilde{c}_{k}]=[\tilde{c}_{0},\tilde{c}_{k}^{\ast
}]=0,~[\tilde{c}_{j},\tilde{c}_{k}^{\ast }]=\delta _{jk}$ for $j,k= 1,\ldots,n_{\mathrm{min}}-1$. By (\ref{eq:b_2}), 
\begin{equation}\label{eq:L_new2}
\tilde{L}=\sqrt{\bar{\gamma}}\tilde{c}_0.
\end{equation}
Define matrices 
\begin{equation}
Q=\left[ 
\begin{array}{ccc}
\tilde{P}_{0}(\bar{\omega}_{1}) & \cdots & \tilde{P}_{0}(\bar{\omega}_{n_{\mathrm{min}}}) \\ 
\vdots & \ddots & \vdots \\ 
\tilde{P}_{n_{\mathrm{min}}-1}(\bar{\omega}_{1}) & \cdots & \tilde{P}_{n_{\mathrm{min}}-1}(\bar{\omega}_{n_{\mathrm{min}}})
\end{array}
\right] \equiv \left[ 
\begin{array}{c}
\tilde{P}_{0}(\bar{\omega}) \\ 
\vdots \\ 
\tilde{P}_{n_{\mathrm{min}}-1}(\bar{\omega})
\end{array}
\right]  \label{eq:Q_temp}
\end{equation}
and $\Gamma \equiv \mathrm{diag}\left( \sqrt{\frac{\bar{\gamma}_{1}}{\bar{\gamma}}},\cdots ,\sqrt{\frac{\bar{\gamma}_{n_{\mathrm{min}}}}{\bar{\gamma}}}\right) $. It can be shown that 
\begin{equation}
\sum_{k=0}^{n_{\mathrm{min}}-1}\frac{\bar{\gamma}_{i}}{\bar{\gamma}}\tilde{P}_{k}(\bar{\omega}_{i})\tilde{P}_{k}(\bar{\omega}_{j})=\delta
_{ij},~~i,j=1,\ldots ,n_{\mathrm{min}},  \label{eq:1.1.14_WG}
\end{equation}
see, e.g., \cite[Eq. (1.1.14)]{WG}. By (\ref{eq:1.1.14_WG}), it can be
verified that the inverse matrix of the matrix $Q$ turns out to be 
$Q^{-1}=\Gamma ^{2}[\tilde{P}_{0}(\bar{\omega})^{\dag }~~\ldots ,~~\tilde{P}_{n_{\mathrm{min}}-1}(\bar{\omega})^{\dag }]$. Thus we have 
\begin{equation}
\left[ 
\begin{array}{c}
\tilde{c}_{0} \\ 
\tilde{c}_{1} \\ 
\vdots \\ 
\tilde{c}_{n_{\mathrm{min}}-1}
\end{array}
\right] =Q\Gamma \left[ 
\begin{array}{c}
\bar{a}_{1} \\ 
\bar{a}_{2} \\ 
\vdots \\ 
\bar{a}_{n_{\mathrm{min}}}
\end{array}
\right] .
\end{equation}
With this, the Hamiltonian of the minimal realization can be re-written as 
\begin{equation} \label{eq:Hamiltonian_temp}
\sum_{j=1}^{n_{\mathrm{min}}}\bar{\omega}_{j}\bar{a}_{j}^{\ast }\bar{a}_{j}=
\left[ 
\begin{array}{c}
\tilde{c}_{0} \\ 
\tilde{c}_{1} \\ 
\vdots \\ 
\tilde{c}_{n_{\mathrm{min}}-1}
\end{array}
\right] ^{\dag }((Q\Gamma )^{-1})^{\dag }\Gamma \left[ 
\begin{array}{ccc}
\bar{\omega}_{1} &  &  \\ 
& \ddots &  \\ 
&  & \bar{\omega}_{n_{\mathrm{min}}}
\end{array}
\right] (Q\Gamma )^{-1}\left[ 
\begin{array}{c}
\tilde{c}_{0} \\ 
\tilde{c}_{1} \\ 
\vdots \\ 
\tilde{c}_{n_{\mathrm{min}}-1}
\end{array}
\right].  
\end{equation}
Finally, according to (\ref{eq:Q_temp}) and (\ref{eq:1.1.14_WG}), we have  the new Hamiltonian matrix
\begin{eqnarray}
\tilde{H}
&=&((Q\Gamma )^{-1})^{\dag }\Gamma \left[ 
\begin{array}{ccc}
\bar{\omega}_{1} &  & 0 \\ 
& \ddots &  \\ 
0 &  & \bar{\omega}_{n_{\mathrm{min}}}
\end{array}
\right] (Q\Gamma )^{-1}  \label{eq:identity_relation} \\
&=&\left[ 
\begin{array}{cccccc}
\tilde{\omega}_{0} & \sqrt{\tilde{\kappa}_{1}} & 0 & \cdots & 0 & 0 \\ 
\sqrt{\tilde{\kappa}_{1}} & \tilde{\omega}_{1} & \sqrt{\tilde{\kappa}_{2}} & 
& 0 & 0 \\ 
0 & \sqrt{\tilde{\kappa}_{2}} & \tilde{\omega}_{2} & \ddots &  & \vdots \\ 
\vdots &  & \ddots & \ddots & \sqrt{\tilde{\kappa}_{n_{\mathrm{min}}-2}} & 0
\\ 
0 & 0 &  & \sqrt{\tilde{\kappa}_{n_{\mathrm{min}}-2}} & \tilde{\omega}_{n_{\rm min}-2} & \sqrt{\tilde{\kappa}_{n_{\mathrm{min}}-1}} \\ 
0 & 0 & \cdots & 0 & \sqrt{\tilde{\kappa}_{n_{\mathrm{min}}-1}} & \tilde{\omega}_{n_{\mathrm{min}}-1}
\end{array}
\right].  \notag
\end{eqnarray}
With the new coupling operator $\tilde{J}$ defined  (\ref{eq:L_new2}) and new Hamiltonian matrix $\tilde{H}$ defined in  (\ref{eq:identity_relation}), the realization (\ref{eq:L_new})-(\ref{eq:H_new}) can be obtained.
 The proof is completed. 

%%%%%%%%%%%%%%%%%%%%%%%%%%%%%%%
\textit{Proof of Lemma \ref{lem:matrix_rep2}.} We again use induction. The
formula is clearly true for $n=1$. Let us set 
\begin{equation*}
N_{n}=\left[ 
\begin{array}{cccc}
a_{0} & b_{1} &  & 0 \\ 
b_{1} & a_{1} & \ddots &  \\ 
& \ddots & \ddots & b_{n} \\ 
0 &  & b_{n} & a_{n}
\end{array}
\right]
\end{equation*}
and so 
\begin{equation*}
N_{n+1}=\left[ 
\begin{array}{cc}
N_{n} & b_{n+1}f_{n} \\ 
b_{n+1}f_{n}^{\top } & a_{n+1}
\end{array}
\right] ,\text{ where }f_{n}=\left[ 
\begin{array}{c}
0 \\ 
\vdots \\ 
0 \\ 
1
\end{array}
\right] \in \mathbb{C}^{n+1}
\end{equation*}
Let us write $E_{11}\left( M\right) $ for the first entry (row 1, column 1) of a matrix $M$. We deduce from the Schur-Feshbach formula (\ref{eq:mar18_appendix_1}) that 
\begin{equation*}
E_{11}\left( N_{n+1}^{-1}\right) =E_{11}\left( (N_{n}-\dfrac{b_{n+1}^{2}}{a_{n+1}}f_{n}f_{n}^{\top })^{-1}\right) .
\end{equation*}
However, the matrix $N_{n}-(b_{n+1}^{2}/a_{n+1})f_{n}f_{n}^{\top }$ is
identical to $N_{n}$ except that we replace the last row, last column entry $a_{n}$ with $a_{n}-(b_{n+1}^{2}/a_{n+1})$, and if by assumption the relation
is true for $n$ we deduce the formula for $n+1$. The formula is true by
induction.

%%%%%%%%%%%%%%%%%%%%%%%%%%%%%%%%
%%%%%%%%%%%%%%%%%%%%%%%%%%%%%%%%
%%%%%%%%%%%%%%%%%%%%%%%%%%%%%%%%

\end{document}